%% file: amain.tex
\documentclass[
letter,
twocolumn,
superscriptaddress,
amsmath,
amssymb,
prl,
nofootinbib
showkeys,
10pt,
floatfix,
nobibnotes,
aps,
fltpage
]{revtex4-2}

\usepackage{latexsym}
\usepackage{amsmath}
\usepackage{amssymb}
\usepackage[T1]{fontenc}
\usepackage[open]{bookmark}
\usepackage{hyperref}
\hypersetup{colorlinks=true,allcolors=blue}
\usepackage{upgreek}

\usepackage{orcidlink}

\usepackage{dcolumn}
\usepackage{bm}
\usepackage{xcolor}
\usepackage{hyperref}
\usepackage{cleveref}
\usepackage{float}
\usepackage{svg}
\usepackage{amssymb}
\usepackage{scalerel}
\usepackage{caption}
\usepackage{wasysym}
\usepackage{lipsum}
\usepackage{enumitem}

\providecommand{\pnl}[1]{{\textcolor{blue}{#1}}}

\makeatletter
\pretocmd\frontmatter@keys@format{\addvspace{20\p@}}{}{}
\makeatother

\usepackage{svg}
\usepackage{cleveref}
\usepackage{siunitx}
\usepackage{caption}
\usepackage{multirow}
\usepackage{makecell}
\usepackage{tabularray}
\usepackage{makecell}  
\usepackage{array}     
\usepackage{graphicx}  
\usepackage{booktabs} 

\begin{document}

\title{An atlas of photonic and plasmonic materials for cathodoluminescence microscopy}

\author{Sven Ebel\,\orcidlink{0009-0005-3224-6413}}
\affiliation{POLIMA---Center for Polariton-driven Light--Matter Interactions, University of Southern Denmark, Campusvej 55, DK-5230 Odense M, Denmark}

\author{Yonas Lebsir\,\orcidlink{0000-0002-9383-0278}}
\affiliation{POLIMA---Center for Polariton-driven Light--Matter Interactions, University of Southern Denmark, Campusvej 55, DK-5230 Odense M, Denmark}

\author{Torgom~Yezekyan\,\orcidlink{0000-0003-2019-2225}}
\affiliation{POLIMA---Center for Polariton-driven Light--Matter Interactions, University of Southern Denmark, Campusvej 55, DK-5230 Odense M, Denmark}

\author{N. Asger Mortensen\,\orcidlink{0000-0001-7936-6264}}
\affiliation{POLIMA---Center for Polariton-driven Light--Matter Interactions, University of Southern Denmark, Campusvej 55, DK-5230 Odense M, Denmark}
\affiliation{Danish Institute for Advanced Study, University of Southern Denmark, Campusvej 55, DK-5230 Odense M, Denmark}

\author{Sergii Morozov\,\orcidlink{0000-0002-5415-326X}}
\affiliation{POLIMA---Center for Polariton-driven Light--Matter Interactions, University of Southern Denmark, Campusvej 55, DK-5230 Odense M, Denmark}

\date{\today}

\begin{abstract}
\vspace{0.0cm}
\textbf{Abstract.} 
Cathodoluminescence (CL) microscopy has emerged as a powerful tool for investigating the optical properties of materials at the nanoscale, offering unique insights into the behavior of photonic and plasmonic materials under electron excitation.
We introduce an atlas of bulk CL spectra and intensity for a broad range of materials used in photonics and plasmonics. 
Through a combination of experimental CL microscopy and Monte Carlo simulations, we characterize spectra and intensity of coherent and incoherent CL, electron penetration depth and energy deposition, offering a foundational reference for interpreting CL signals and understanding material behavior under electron excitation. 
Our atlas captures CL signals across a wide range of materials, offering valuable insight into intrinsic emission properties for informed material selection and device design in photonics and plasmonics.
\vspace{1.0cm}
\end{abstract}


\maketitle

\section{1 Introduction}

Cathodoluminescence (CL) microscopy is an indispensable tool for investigating the optical and electronic properties of materials at the nanoscale, providing valuable insights into photonic and plasmonic materials~\cite{Yacobi1986,Coenen2016,Coenen2017,Christopher2020,RoquesCarmes2023}. 
When an electron beam interacts with a material, it can generate a broad range of emissions, including X-rays, secondary and backscattered electrons, and photons spanning from the ultraviolet (UV) to the infrared (IR) spectral range. 
Among these, UV to IR photons are of particular interest for photonics and plasmonics.
The detection of CL optical signal combined with the techniques of electron microscopy allows for the spatial mapping of the material’s electronic and optical characteristics with nanoscale resolution. 
The wavelength and intensity of the emitted light are influenced by the material’s composition, defects, and structural features, making CL microscopy a valuable technique for understanding material properties and behavior.
For in-depth insights into various aspects of CL, readers are directed to several comprehensive reviews, which include discussions on the physical fundamentals of CL~\cite{GarciaDeAbajo2010,GarcadeAbajo2021}, its historical development and methods~\cite{Brillson2012,Coenen2016,Coenen2017,Kociak2017,Meuret2019,Christopher2020,Yoshikawa2023,Dang2023}, overviews on CL origin in a variety of semiconductors~\cite{Yacobi1986,Yacobi1990,Guthrey2020}, and developments on CL microscopy for the study of nanophotonic and plasmonic systems~\cite{Polman2019,GmezMedina2008,RoquesCarmes2023}. 

In combination with far-field Fourier spectroscopy, CL microscopy offers a powerful method for distinguishing between coherent and incoherent light emission processes in materials~\cite{Coenen2011,Brenny2014,Mignuzzi2018,Scheucher2022,Fiedler2022dis}. 
Coherent CL arises from transition radiation, Smith--Purcell radiation, and collective oscillations of free electrons, such as surface-plasmon resonances in metals, while incoherent CL results from intrinsic processes (e.g., electron-hole recombination, defects) and extrinsic processes (e.g., dopants, impurity and color centers), which are typically observed in semiconductors and dielectrics~\cite{Brenny2014,Yanagimoto2025}.
Differentiating and accurately attributing these mechanisms is important in photonic and plasmonic systems, as they often involve materials that support both processes. 
The relative strength of incoherent and coherent mechanisms strongly depends on the system's material composition. 
Therefore, for effective experiment design and interpretation, it is essential to understand the CL response of the materials used, independent of geometric effects, and to identify the origins of the observed spectral features.

An important parameter in CL microscopy is the electron penetration depth, which refers to how deeply the electron beam penetrates the material before losing its kinetic energy. 
The penetration depth varies depending on the material’s atomic number ($Z$), atom density, and the energy of electrons in the electron beam~\cite{Drouin2007}.
For low-$Z$ materials like carbon (C) or silicon (Si), penetration depths can be relatively large (on scale of microns), allowing for bulk material analysis even at lower electron beam voltages. 
Conversely, in high-$Z$ materials like gold (Au) or platinum (Pt), the penetration depth is significantly reduced (down to sub 10\,nm), restricting the interaction volume to the surface or near-surface layers. 
Understanding the electron penetration depth is relevant for interpreting CL accurately, as it determines whether the emission originates from the surface, subsurface, or bulk of the material~\cite{Norris1973,Tizei2013,Francaviglia2022}.
The penetration depth and corresponding distribution of absorbed energy plays a key role in electron beam lithography, where precise energy deposition is essential for controlling pattern resolution and resist exposure~\cite{Drouin2007}.

\begin{figure}[!htbp]
\includegraphics[width=0.99\linewidth]{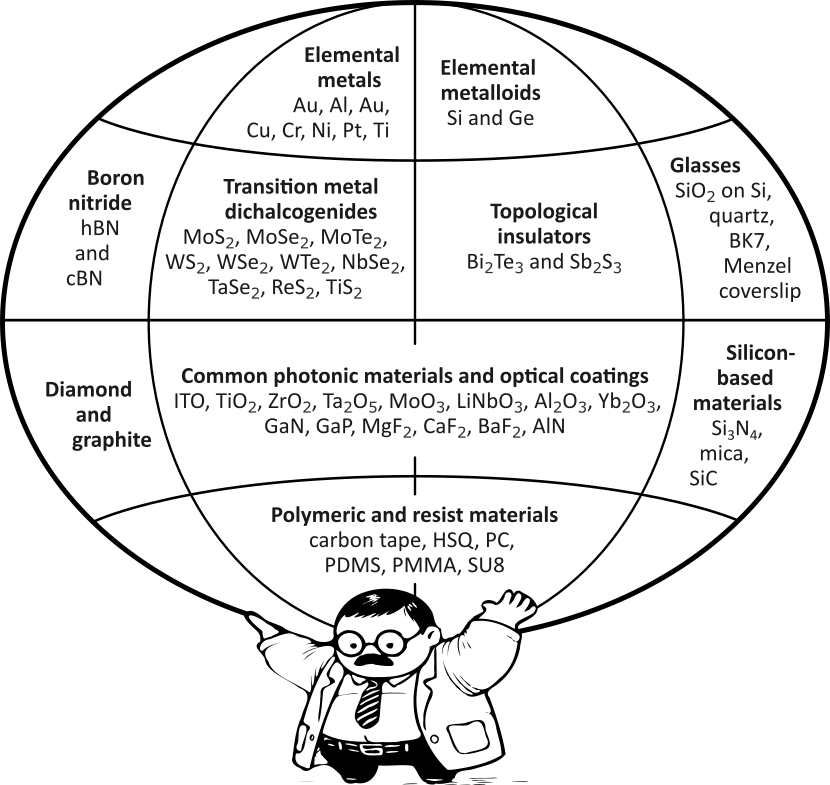}
\caption{\textbf{An atlas of investigated photonic and plasmonic materials. }
}
\label{fig-materials}
\end{figure}

\begin{figure*}[!htbp]
\includegraphics[width=0.95\linewidth]{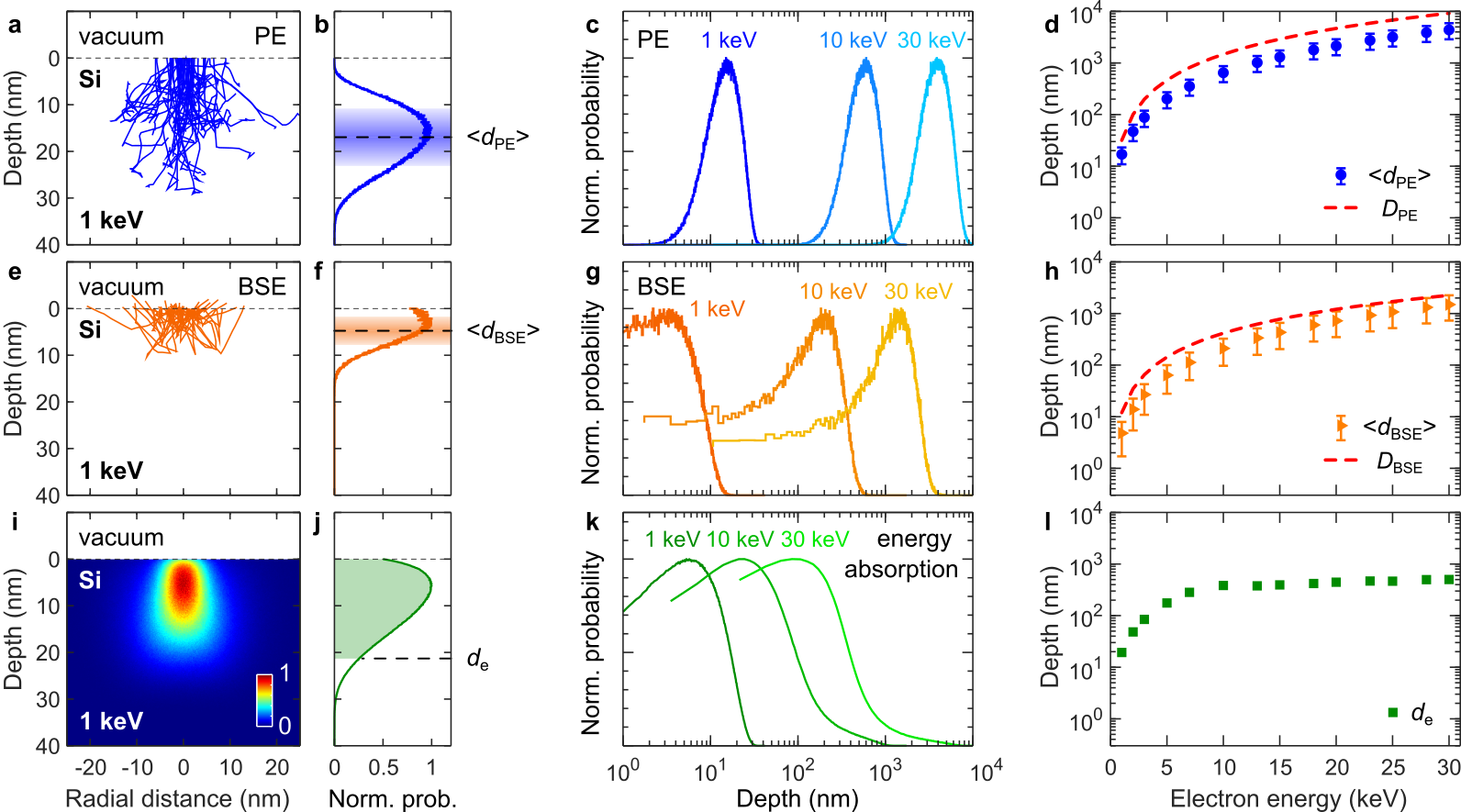}
\caption{\textbf{Monte Carlo simulations of electron trajectories and energy deposition in bulk Si at increasing electron beam energy.}
\textbf{a,e}~PE and BSE trajectories for an incident electron beam at 1\,keV with 10\,nm diameter (shown only 50~trajectories).
Corresponding distributions of the PE (\textbf{b}) and BSE (\textbf{f}) depth at 1\,keV~(acquired for $10^6$ incident electrons). The horizontal dashed lines represent the weighted mean $\langle d_\mathrm{PE}\rangle$ and $\langle d_\mathrm{BSE}\rangle$, while the shaded areas show two standard deviations.
Distributions of the PE (\textbf{c}) and BSE (\textbf{g}) depth at increasing electron beam energy~(acquired for $10^6$ incident electrons).
\textbf{d,h}~Increase of $\langle d_\mathrm{PE}\rangle$ and $\langle d_\mathrm{BSE}\rangle$ with electron energy. The error bars represent two standard deviations as shown in the corresponding panels \textbf{b} and \textbf{f}. 
The dashed red lines are the Kanaya--Okayama range $D_\mathrm{PE}$ for PE in \textbf{d}, and BSE origin depth $D_\mathrm{BSE}$ in \textbf{h}.
\textbf{i}~Energy deposition map at 1\,keV.
\textbf{j}~Axial profile of the energy deposition map from panel~\textbf{i}. The dashed line indicates the 90\% energy loss  depth $d_\mathrm{e}$.
\textbf{k}~Axial profiles of the energy deposition at increasing electron beam energy.
\textbf{l}~Increase of $d_\mathrm{e}$ with electron beam energy.
}
\label{fig-simulations}
\end{figure*}

In this article, we present a comprehensive study of the optical response from a wide range of materials under low-energy electron beam excitation (1--30\,keV) in a scanning electron microscope (SEM), building up an atlas of CL spectra of materials for photonic and plasmonic systems. 
The materials studied include plasmonic and non-plasmonic metals, dielectrics, and semiconductors used in optics and photonics, as well as novel two-dimensional (2D) materials, topological insulators, and polymeric and resist materials used in the fabrication of nanostructured devices.
Our aim is to probe the CL emission mechanisms (coherent/incoherent) and to characterize the emitted spectra and CL intensity from each material.
We also perform numerical simulations of electron penetration depth and energy absorption using Monte Carlo methods via the CASINO modeling tool~\cite{Drouin2007}. 
These simulations provide detailed insights into the interaction of low-energy electrons with the investigated materials. 
Our work combines experimental CL analysis with Monte Carlo simulations, offering a robust dataset to understand the impact of material properties and electron interactions on CL behavior.
Furthermore, this atlas serves as a reference for interpreting CL data in various experimental contexts, reducing the likelihood of misinterpretation of emission origins, material properties, or electron-matter interactions. 
By systematically documenting the CL characteristics of key materials, this work facilitates the development of next-generation photonic and plasmonic technologies, guiding both experimental design and theoretical modeling efforts.

\section{2 Materials}
We present CL properties of a diverse range of commercially available materials (Fig.~\ref{fig-materials}), which serve as potential building blocks and platforms for emerging photonic and plasmonic devices. 
Each material is examined using low-energy (1--30\,keV) and low current (ca. 1.4\,nA) electron beams, allowing us to characterize the origin of CL emissions -- whether coherent or incoherent -- and analyze their spectra and intensity. 
Furthermore, the Supplementary Materials (SM) provide detailed information on the investigated materials, including their suppliers and origins, Raman spectra, and morphological characterization using optical and electron-based imaging techniques, as well as simulation results of electron beam interaction with the materials.

\begin{figure*}[!htbp]
\includegraphics[width=0.95\linewidth]{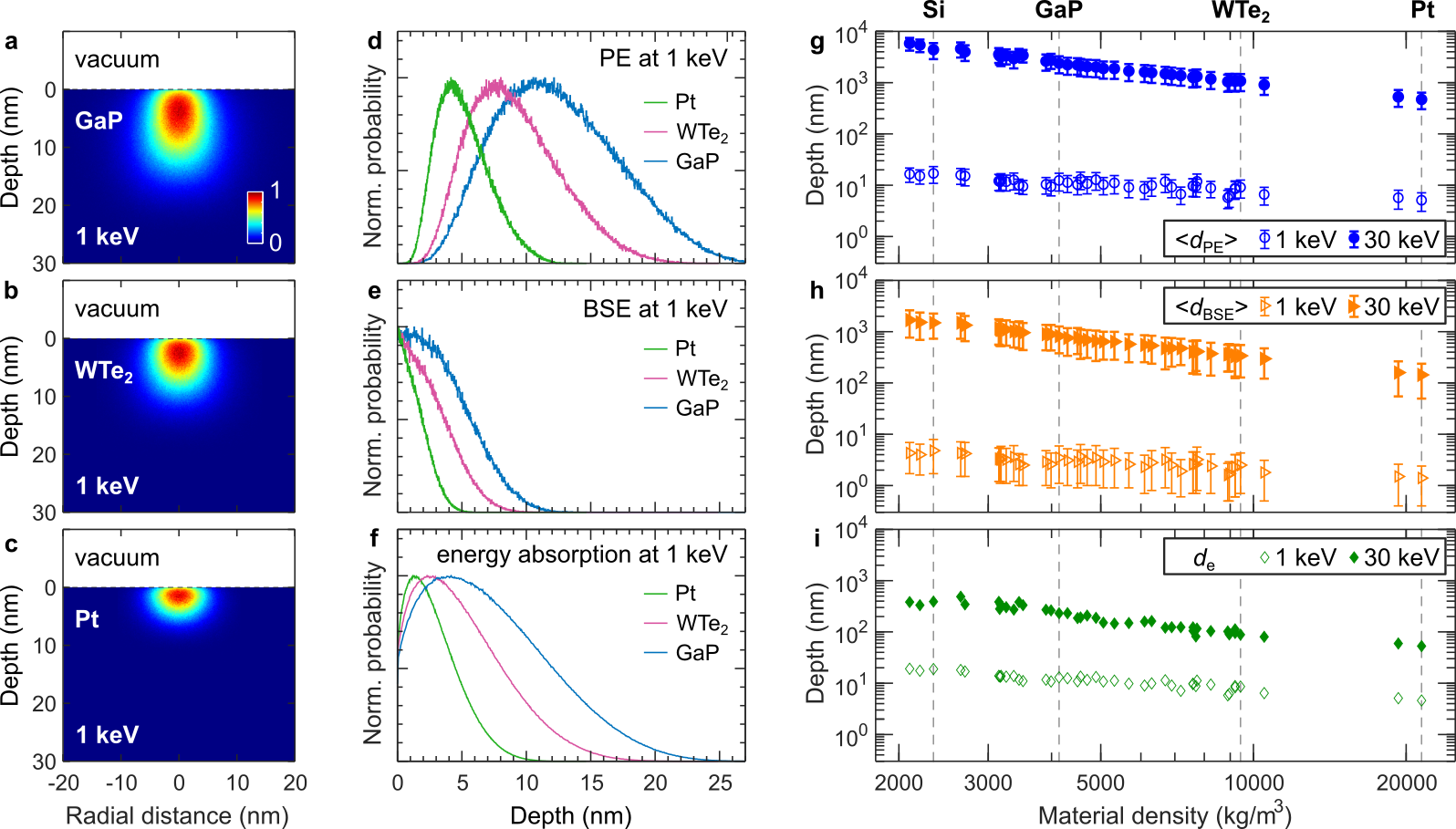}
\caption{\textbf{Monte Carlo simulations of electron propagation and energy deposition in a variety of bulk materials.}
\textbf{a-c}~Normalized spatial maps of 1\,keV electron beam energy deposition in GaP, WTe$_2$, and Pt. 
\textbf{d-f}~1\,keV distributions of $d_\mathrm{PE}$, $d_\mathrm{BSE}$, and axial profiles of the absorbed power in GaP, WTe$_2$ and Pt.
\textbf{g-i}~Decrease of $\langle d_\mathrm{PE}\rangle$, $\langle d_\mathrm{BSE}\rangle$, and $d_\mathrm{e}$ for denser bulk materials at 1\,keV~(open symbols) and 30\,keV~(filled symbols). 
Densities of Si, GaP, WTe$_2$, and Pt are highlighted with vertical dashed lines. The error bars represent two standard deviations. 
}
\label{fig-simulations-materials}
\end{figure*}

\section{3 Monte Carlo simulations}
We use Monte Carlo simulations performed with CASINO v2.42~\cite{Drouin2007} to model electron trajectories and the distribution of absorbed energy in various bulk materials under electron-beam excitation.
CASINO software employs stochastic methods to simulate multiple electron scattering events, energy loss mechanisms, and backscattering, providing a spatial map of electron trajectories for primary electron (PE) and backscattered electrons (BSE). 
CASINO simulates elastic scattering and approximates inelastic events using the mean energy loss between successive elastic scattering interactions~\cite{Hovington1997p3,Hovington1997p1}. 
All our simulations utilize interpolation from the Mott cross-section database~\cite{Drouin1997p2}, while the Joy--Luo stopping power model~\cite{Joy1989} is applied to account for ionization potential, collisions, and inelastic energy loss~\cite{Hovington1997p3}. 
For a more in-depth discussion on the theory of free electron–matter interactions, which underpin the Monte Carlo simulations used in our work, we refer the reader to the books by Reimer~\cite{Reimer1998} and Egerton~\cite{Egerton2011}.

We begin by presenting simulation results in Fig.~\ref{fig-simulations} on electron beam interaction with silicon (Si), exploring how varying electron beam energy influences electron propagation and energy deposition.
Fig.~\ref{fig-simulations}\pnl{a} presents the PE trajectory maps at 1\,keV. 
In total, we simulate $10^6$ electrons, though only 50 trajectories are shown here for clarity. 
Fig.~\ref{fig-simulations}\pnl{b} shows the corresponding statistics for the maximum penetration depth $d_{\mathrm{PE}}$ -- defined as the point where electron energy falls below 0.1\,keV.
Higher electron beam energy results in deeper PE distributions, as shown in Fig.~\ref{fig-simulations}\pnl{c}.
We quantify these results by calculating the weighted mean of PE depth $\langle d_\mathrm{PE}\rangle = {\sum p_i d_{\mathrm{PE,}i}}$ where $p_i$ represents the normalized probabilities, and two standard deviations, as illustrated in Fig.~\ref{fig-simulations}\pnl{b}. 
Fig.~\ref{fig-simulations}\pnl{d} summarizes the obtained $\langle d_\mathrm{PE}\rangle$ in Si over the electron beam energy range of 1--30\,keV. 
As the electron beam energy increases, $\langle d_\mathrm{PE} \rangle$ extends deeper into Si, from approximately 16\,nm at 1\,keV to nearly 5\,$\upmu$m at 30\,keV. 
We compare these results to the Kanaya--Okayama range $D_\mathrm{PE}$, which provides an estimate of how deep an electron can travel in a material before being fully absorbed, primarily considering energy loss due to inelastic scattering~\cite{KKanaya1972}.
The range is given by $D_\mathrm{PE} \text{ [nm]}= 27.61\cdot 10^{-5}A\cdot E^{5/3}/(\rho \cdot Z^{8/9})$, where 27.61 is a dimensional consistency factor, $A$ is the atomic weight in g/mol, $E$ is the electron energy in eV, $\rho$ is the material density in g/cm$^3$, and $Z$ is the material’s atomic number. 
We plot $D_\mathrm{PE}$ as a red dashed line alongside the Monte Carlo simulation results for $\langle d_\mathrm{PE} \rangle$ in Fig.~\ref{fig-simulations}\pnl{d}. 
We note that the PE penetration depth increases with electron energy, following a power law trend.

In CASINO simulations, BSEs are defined as PEs that escape the sample surface after multiple scattering events, provided their energy remains above 0.1\,keV upon exiting the material.
Fig.~\ref{fig-simulations}\pnl{e} presents the BSE trajectory maps in Si at 1\,keV, and Fig.~\ref{fig-simulations}\pnl{f} shows the corresponding statistics for the BSE origin depth $d_\mathrm{BSE}$, which is defined as the depth of the first large-angle elastic scattering event (greater than 90$^\circ$ deflection) that initiates the trajectory leading the electron back to the surface. 
Higher electron beam energy results in deeper BSE origin depth distribution, as shown in Fig.~\ref{fig-simulations}\pnl{g}.
Similarly to PE, we characterize the BSE origin depth with weighted mean $\langle d_\mathrm{BSE}\rangle$ and two standard deviations, as illustrated in Fig.~\ref{fig-simulations}\pnl{f}.
Fig.~\ref{fig-simulations}\pnl{h} summarizes the obtained $\langle d_\mathrm{BSE}\rangle$ in Si over the electron beam energy range of 1--30\,keV. 
$\langle d_\mathrm{BSE}\rangle$ increases in Si with electron energy from about 5\,nm at 1\,keV to 1.5\,$\upmu$m at 30\,keV. 
We compare these results with experimental observations of BSE range~\cite{Reimer1977,Reimer1998}, which is described by the empirical expression $D_\mathrm{BSE}\text{ [nm]} = 28\cdot E^{1.54}/\rho$, where 28 is an experimental scaling factor, $E$ is the electron energy in keV, and $\rho$ is the material density in g/cm$^3$.
We plot $D_\mathrm{BSE}$ as a red dashed line alongside the Monte Carlo simulation results for $\langle d_\mathrm{BSE}\rangle$ in Fig.~\ref{fig-simulations}\pnl{h}, observing a power law trend. 

We also investigate the impact of increasing electron beam energy on the spatial distribution of absorbed energy, as it is relevant for determining the origin of the CL signal.
Fig.~\ref{fig-simulations}\pnl{i} shows the spatial profile of electron energy deposition in Si at 1\,keV, with the total energy summed for each depth in Fig.~\ref{fig-simulations}\pnl{j}.
We extract the depth $d_\mathrm{e}$ at which incident electrons lose 90\% of their energy from the energy deposition profile, as illustrated in Fig.~\ref{fig-simulations}\pnl{j}.
Furthermore, Fig.~\ref{fig-simulations}\pnl{k} presents the evolution of energy absorption distributions for 1\,keV, 10\,keV, and 30\,keV.
Fig.~\ref{fig-simulations}\pnl{l} summarizes the increase of $d_\mathrm{e}$ in Si with electron beam energy. 
Understanding the absorbed energy profiles helps in determining the effective electron-matter interaction volume, which is crucial for designing of CL experiments, CL characterization of photonic and plasmonic devices, and optimizing electron-beam lithography processes.

Taking a broader perspective, Fig.~\ref{fig-simulations-materials} investigates how material density affects the characteristic depths $\langle d_\mathrm{PE} \rangle$, $\langle d_\mathrm{BSE} \rangle$ and $d_\mathrm{e}$. 
Fig.~\ref{fig-simulations-materials}\pnl{a-c} illustrate the reduction in electron-matter interaction volume as material density increases.
To highlight these effects, we selected gallium phosphide (GaP, $\rho$=4,138\,kg/m$^3$), tungsten ditelluride (WTe$_2$, $\rho$=9,430\,kg/m$^3$), and platinum (Pt, $\rho$=21,450\,kg/m$^3$) for their distinctive densities. 
This selection emphasizes the differences in characteristic depth distributions of PE, BSE, and energy absorption at 1\,keV, as shown in Fig.~\ref{fig-simulations-materials}\pnl{d-f}.
We repeat these simulations for a wide range of materials with densities spanning from 2,100\,kg$/$m$^{3}$ for boron nitride (BN) to 21,450\,kg$/$m$^{3}$ for Pt and plot the characteristic depths at 1\,keV and 30\,keV in Fig.~\ref{fig-simulations-materials}\pnl{g-i}. 
The general trend indicates that characteristic electron depths decrease as material density increases. 
This effect is more pronounced at the higher electron beam energy of 30\,keV (note the steeper slope for filled symbol dataset in Fig.~\ref{fig-simulations-materials}\pnl{g-i}). 

\begin{figure}[b]
\includegraphics[width=0.95\linewidth]{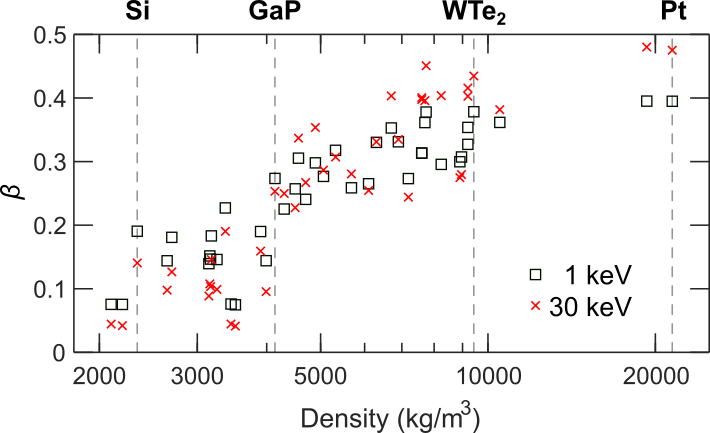}
\caption{\textbf{Monte Carlo simulations of backscattering coefficient $\beta$ in bulk materials.}
Increase of $\beta$ with material density at 1\,keV (squares) and 30\,keV (crosses). 
Densities of the selected materials are indicated by vertical dashed lines.
}
\label{fig-simulations-beta}
\end{figure}

Finally, Fig.~\ref{fig-simulations-beta} summarizes the results of Monte Carlo simulations for backscattering coefficient $\beta$, which is the ratio of BSE to incident PE and provides an estimate of the electrons involved in inelastic interactions responsible for generating CL. 
A higher backscattering coefficient reduces the effective beam current available for generating CL, making it an important factor in evaluating CL efficiency.
Interestingly, Fig.~\ref{fig-simulations-beta} demonstrates that $\beta$ is higher for low-density materials like Si at low electron beam energy of 1\,keV, while the trend reverses for high-density materials like Pt. 
This result has been previously observed experimentally as a consequence of the interplay between elastic and inelastic scattering, where low-energy electrons in low-density materials penetrate deeper with less scattering, leading to higher $\beta$, while high-density materials exhibit stronger elastic scattering at higher energies, increasing 
$\beta$~\cite{Darlington1972}.
For reference, we provide a detailed listing in the SM of $\langle d_\mathrm{PE}\rangle$, $\langle d_\mathrm{BSE}\rangle$, $d_\mathrm{e}$, and $\beta$ for all the investigated materials.

Having modeled electron–material interactions through Monte Carlo simulations, we gained detailed insight into how electron beam energy and material properties influence penetration depth, energy deposition, and backscattering behavior. 
Low electron beam energies primarily excite near-surface regions, while high energies penetrate deeper, exciting a larger volume; in low-density materials, excitation occurs gradually over greater depths, whereas in high-density materials, energy is deposited closer to the surface due to stronger scattering.
Building on these simulation results, the following sections outline the experimental procedures used to acquire and interpret CL emission data.

\begin{figure*}[t]

\includegraphics[width=0.99\linewidth]{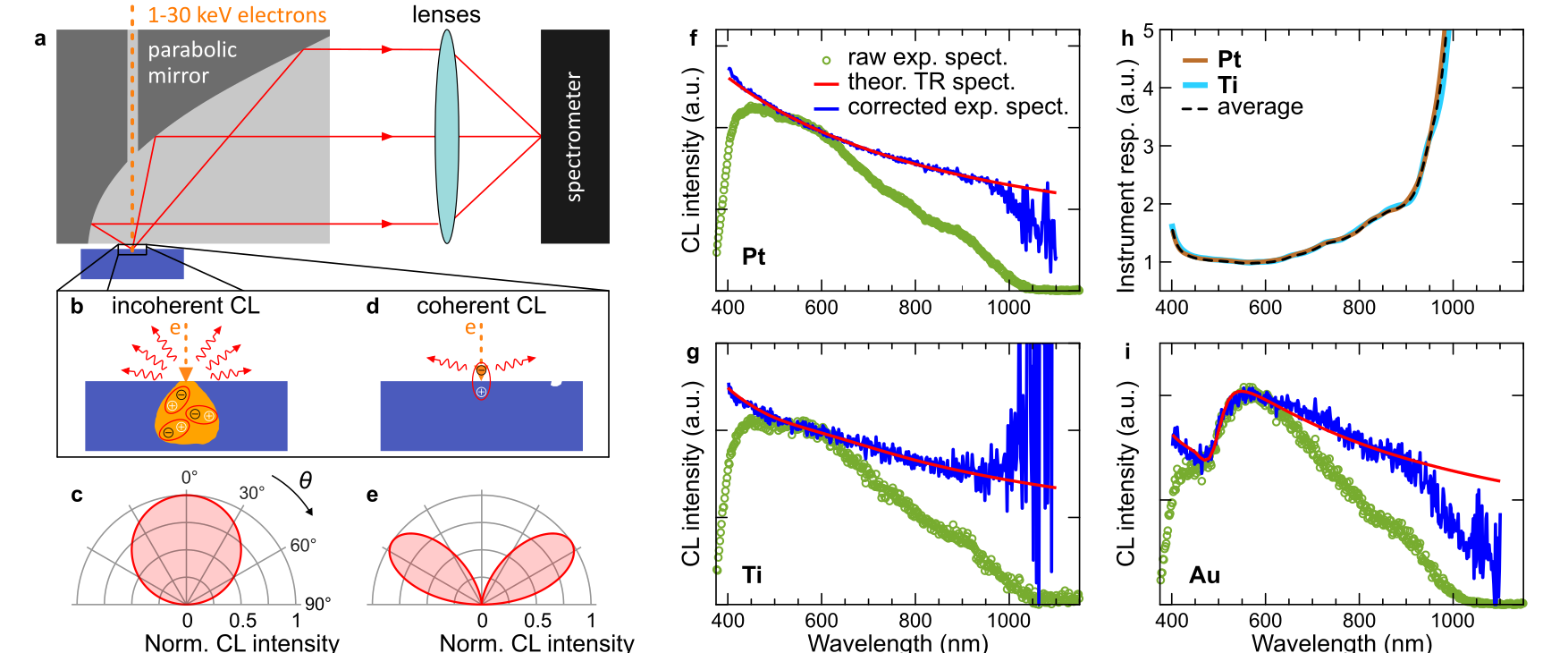}
\caption{\textbf{Cathodoluminescence microscopy.}	
\textbf{a}~Experimental setup for detection of CL emission and angular spectra.
\textbf{b}~Incoherent CL originates from the recombination of electron-hole pairs generated by incident electrons in the bulk and is characterized by an isotropic angular radiation pattern (\textbf{c}).
\textbf{d}~Coherent CL arises from the interaction of an electron beam with collective electron oscillations, as for example transition radiation (TR), and is characterized by a dipolar emission pattern (\textbf{e}). 
\textbf{f-g}~Experimental TR spectra of Pt and Ti at 30\,keV (green circles) are used to calibrate instrument response by comparing them with theoretical TR spectra (red), and obtaining the corrected CL spectra (blue, see main text for details).
\textbf{h}~Instrument response functions obtained with TR of Pt (brown) and Ti (cyan), and their average (dashed black).
\textbf{i}~The averaged instrument response function is applied to the experimental CL spectrum of gold at 30\,keV (green circles), producing a corrected spectrum (blue), that closely matches the theoretical TR spectrum of gold (red). 
}
\label{fig-intro}
\end{figure*}

\section{4 Experimental methods}
\subsection{4.1 Experimental setup}
The experimental setup for CL microscopy (SPARC Spectral, Delmic) involves a scanning-electron microscope (SEM, TESCAN MIRA3) equipped with a parabolic mirror (effective NA=0.97) to collect the generated emission~(Fig.~\ref{fig-intro}\pnl{a}).
The electron beam passes through a small aperture in the parabolic mirror, interacting with the sample in its focal spot. 
The generated CL is collected by the same mirror and directed towards a spectrometer ({Kymera-193i with Newton~920P, Andor}). 
Our CL spectral measurements were performed at room temperature using a $1\times1$~$\mu$m$^2$ spot size with fuzzing (scanning within the pixel during signal acquisition); the integration time was adjusted based on the brightness of the CL signal, and dark counts were subtracted from all spectra.
For angular-resolved measurements, a set of interchangeable lenses is used to image the Fourier plane, enabling the retrieval of angular radiation profiles and providing insight into the coherent or incoherent origin of the CL signal~\cite{Coenen2011, Brenny2014}.

\subsection{4.2 Angular emission}
Due to the uncorrelated nature of incoherent CL processes (Fig.~\ref{fig-intro}\pnl{b}), the emitted light has an isotropic or Lambertian emission profile, where the intensity is uniform across all directions and decreases with the cosine of the emission angle~(Fig.~\ref{fig-intro}\pnl{c}). 
In contrast, the coherent CL originates from ordered, collective oscillations of electrons~(Fig.~\ref{fig-intro}\pnl{d}).
These coherent processes result in a dipolar emission pattern, where the intensity is strongest along the interface, typically aligned with the orientation of the dipole or the surface-plasmon mode.
This angular dependence is due to the phase coherence of the emitted light, which causes constructive interference in specific directions, resulting in a non-uniform emission profile. 
Due to the presence of the interface between vacuum and sample, the intensity $I(\theta)$ of the dipolar emission pattern is bent towards the normal to interface, and can be modeled as $I(\theta)\propto\sin^2(\theta) \cos(\theta)$~(Fig.~\ref{fig-intro}\pnl{e}), where $\theta$ is an angle from the normal (see Fig.~\ref{fig-intro}\pnl{c}). 
We note that all reported experimental angular radiation profiles are collected within a $\theta$ range of $7^\circ$ to $78^\circ$, with a dip around $\theta = 0^\circ$ caused by the central hole in the parabolic mirror for the electron beam, and truncation at large angles due to the mirror’s limited numerical aperture.

The angular radiation pattern of CL provides insight into the specific radiation mechanism responsible for the emitted light.
Furthermore, the intensity of coherent CL is strongly dependent on the electron beam energy and is generally dimmer compared to its incoherent counterpart~\cite{Brenny2014}. 
In addition, coherent CL is broadband, whereas incoherent CL can exhibit very narrow emission lines, even at room temperature, as seen in color centers in diamonds~\cite{Fiedler2023}.
We also observe that certain materials exhibit either coherent or incoherent CL, whereas others display a mixture of both or can transition from incoherent to coherent CL as the electron beam energy increases.

\subsection{4.3 Instrument response}

In CL spectroscopy, accurate broadband spectral measurements require correction for the instrument’s response function, which accounts for wavelength-dependent variations in detection efficiency~\cite{Brenny2014}. 
Several factors can introduce spectral distortions in our measurements. 
The spectrometer grating used (150 lines/mm, 500\,nm blaze) has its highest diffraction efficiency around 500--600\,nm, meaning wavelengths outside this range may be underrepresented. 
Additionally, the silicon-based photodetector exhibits a significant drop in sensitivity below ca. 400\,nm and beyond 950\,nm, affecting signal strength in these spectral regions. 
Optical components in the CL system, such as the fused silica SEM window and focusing optics, have spectrally dependent transmittance, introducing further variations across the detection range.

To account for these potential distortions and ensure reliable spectral measurements, we apply an instrument response correction, which is particularly crucial for broadband emission sources like transition radiation.
Ensuring accurate spectral calibration allows for a meaningful comparison between experimental data and theoretical models, which predict transition radiation based on the material's optical properties and the electron beam parameters. 
In our study, transition radiation is modeled using the formalism described in prior work by Garc{\'i}a de Abajo~\cite{GarciaDeAbajo2010}, where Maxwell’s equations are solved for a fast electron interacting with a material, inducing surface currents that generate a characteristic transition radiation spectrum~(see details in SM).
This modeling approach incorporates ellipsometry data to determine the dielectric function of the investigated materials, ensuring accurate spectral predictions.

To implement the instrument response correction, we first measure CL spectra of platinum (Pt) and titanium (Ti), as these metals exhibit well-characterized and uniform transition radiation. 
We use two metals to measure the instrument response in order to account for metal-specific variations in transition radiation emission.
Moreover, to minimize geometry-related spectral contribution, experimental CL spectra of Pt and Ti are averaged over multiple positions on the samples.
These experimental CL spectra for Pt and Ti at 30\,keV are shown in Fig.~\ref{fig-intro}\pnl{f,g} (green circles) along with their theoretical transition radiation spectra (red curves).
For calculating the theoretical spectra, we employ ellipsometry data for Pt and Ti from different experimental groups~\cite{Tselin2024,Palm2018}, which helps to further average our instrument response results.
By comparing the experimental spectra with the theoretical ones, we extract instrument response functions for each metal (brown for Pt and cyan for Ti in Fig.~\ref{fig-intro}\pnl{h}).
Here we smoothed the experimental CL spectra for the calculation of the instrument response functions, ensuring a more representative and uniform spectral correction.
We average the two response functions to minimize material-specific variations and systematic errors (dashed black in Fig.~\ref{fig-intro}\pnl{h}).
As expected, the instrument response remains flat in the 500--600\,nm range, where the grating achieves its highest diffraction efficiency, but sharply diverges beyond 900\,nm, resulting from the poor detection efficiency of the silicon-based CCD camera in that region.
Finally, in Fig.~\ref{fig-intro}\pnl{f-g} this averaged instrument response is applied to the raw CL spectra of Pt and Ti (green circles), resulting in the corrected CL spectra (blue).

\begin{figure*}
\includegraphics[width=1\linewidth]{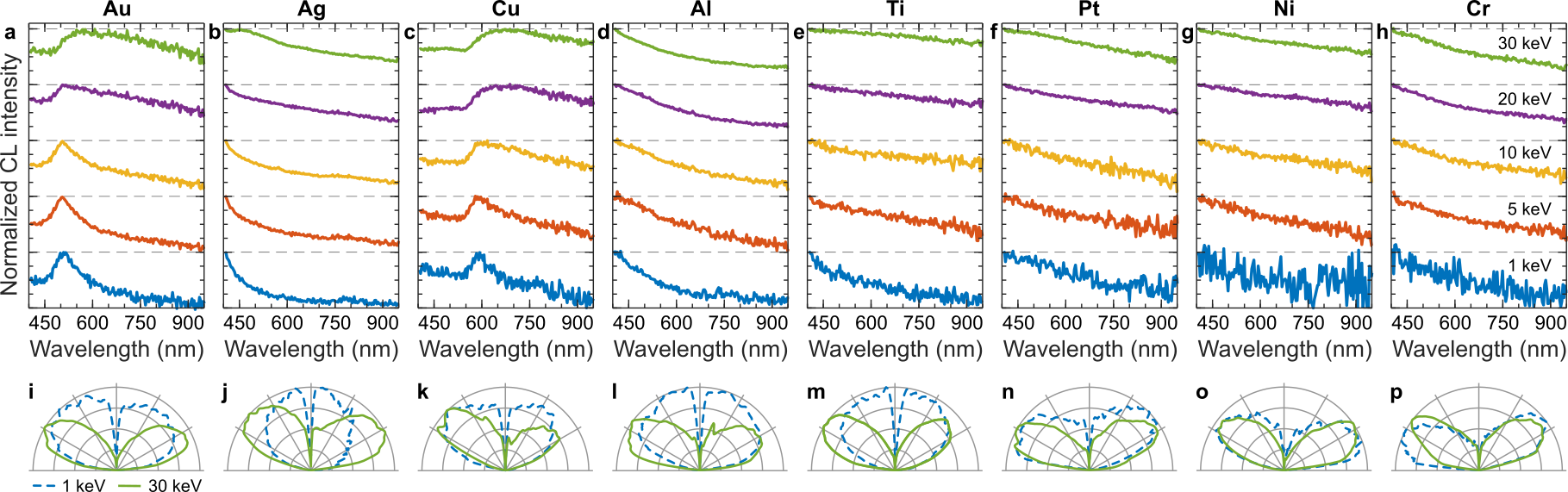}
\caption{\textbf{CL of elemental metals.} 
Normalized CL spectra measured in a range of electron beam energies (see panel \textbf{h} for color coding, the spectra are vertically offset for clarity of presentation) for:
\textbf{a}~mono-crystalline Au, 
\textbf{b}~Ag,
\textbf{c}~Cu,
\textbf{d}~Al,
\textbf{e}~Ti,
\textbf{f}~Pt,
\textbf{g}~Ni,
\textbf{h}~Cr.
\textbf{i-p}~Corresponding angular radiation profiles measured at 1\,keV~(dashed blue) and 30\,keV~(solid green). 
}
\label{fig-spec-emetals}
\end{figure*}

We validate independently the obtained instrument response function using CL of gold (Au), as it has a well-characterized spectrum~\cite{Brenny2014}. 
Fig.~\ref{fig-intro}\pnl{i} presents the experimental spectrum of Au at 30\,keV (green circles) alongside the theoretical transition radiation spectrum (red), calculated using ellipsometry data from {Ref.~\cite{Olmon2012}}. 
The corrected spectrum (blue) is obtained by applying the instrument response function, showing excellent agreement with the theoretical spectrum and accurately reconstructing spectral features in the 400--600\,nm range, attributed to distinct variations in the real and imaginary parts of the Au dielectric function, which influence the transition radiation intensity and spectral shape. 
As expected, the correction becomes unreliable beyond ca.~950\,nm due to the low sensitivity of the silicon detector. 
Consequently, we report corrected spectra within the 400--950\,nm range in our CL atlas, while raw spectra covering 400--1100\,nm are provided in  {SM Fig.~S1-S10}. 
This process ensures that the spectral shape and intensity accurately reflect the intrinsic CL response of the materials reported in our atlas, eliminating distortions introduced by the CL detection system.
Furthermore, another common source of distortion is the second-order diffraction artifact, which arises when bright emission sources produce unwanted spectral contributions at twice their fundamental wavelength. 
To mitigate this effect for materials with bright peaks in the UV region, we used a 450\,nm long-pass filter and report CL spectra up to 900\,nm.

\section{5 Cathodoluminescence spectra}
In this section, we present CL spectra obtained at electron beam voltages from 1\,keV to 30\,keV, demonstrating the effects of different voltages on various photonic and plasmonic materials.
Lower beam voltages ($\leq$5\,kV) probe near-surface regions, revealing surface states and possible contaminants, while higher voltages (30\,kV) penetrate deeper, uncovering bulk material response (Fig.~\ref{fig-simulations} and Fig.~\ref{fig-simulations-materials}). 
This systematic reporting of CL spectra across a range of voltages helps to select appropriate experimental conditions, including background-free substrates for the desired spectral range and optimized beam energy for specific emission characteristics, aiding CL analysis in material science and nanotechnology.

\subsection{5.1 Elemental metals}
CL studies of elemental metals began gaining attention in the mid-20\textsuperscript{th} century, focusing on the optical and electronic properties of metals under electron beam excitation. 
Metals like gold (Au), silver (Ag), copper (Cu), and aluminum (Al) were studied to explore their potential luminescence properties despite their free electron nature, primarily to elucidate the mechanism of transition radiation~\cite{Goldsmith1959,Cram1967,Wartski1975,Papanicolaou1976}.
The 1980s saw a significant increase in interest, particularly due to advancements in CL spectroscopy, which allowed for the study of metal-semiconductor interfaces and the identification of related interface states~\cite{Brillson1985,Brillson1988,Viturro1988,Brillson2012}. 
CL emission from metals could reveal unique interactions between conduction band electrons and d-band holes~\cite{Bashevoy2006}, Mie resonances in metallic spheres~\cite{Yamamoto2001}, as well as surface plasmon effects~\cite{Ferrell1958,Bashevoy2006,vanWijngaarden2006,Yamamoto2006,Vesseur2007,Yamamoto2012,Knight2012,Coenen2014,Schmidt2020,Chi2021,Vu2023,Akerboom2024}.
Today, CL methods are very appealing in nano-optics and plasmonics as they enable high-resolution, nanoscale mapping of modes in metallic nanostructures~\cite{Yamamoto2006,Vesseur2007,GmezMedina2008,Kociak2014,Losquin2015,Day2015,Singh2018,Bittorf2021,Lingstdt2023}. 
Accurate interpretation of CL arising from geometrically-induced plasmonic effects requires understanding the bulk optical response. 

We begin our analysis by presenting CL spectra of bulk metal samples, including the transition metals Au, Ag, Cu, platinum (Pt), nickel (Ni), and chromium (Cr), as well as the post-transition metals such as aluminum (Al) and titanium (Ti). 
Fig.~\ref{fig-spec-emetals}\pnl{a-h} show the CL spectra of these metals across a range of electron beam energies from 1\,keV to 30\,keV. 
At the high electron beam energy of 30\,keV, CL spectra are broadband in all the tested metals, which is characteristic for the transition radiation with the distinctive dipolar, doughnut-shaped, angular radiation profile (green in Fig.~\ref{fig-spec-emetals}\pnl{i-p}).
This occurs because faster electrons interact more strongly with the dense metal structure, generating stronger transition radiation as they cross the vacuum-metal interface~\cite{Piestrup1991,GarciaDeAbajo2010}.

At the lower electron beam energies, the contribution of transition radiation to the CL spectra is significantly reduced~\cite{Chen2023}. 
At 1\,keV, we observe a perfectly isotropic angular radiation profile for Ag (blue in Fig.~\ref{fig-spec-emetals}\pnl{j}) and mixed isotropic-dipolar angular profiles for Au and Cu~(Fig.~\ref{fig-spec-emetals}\pnl{i,k}), which are transition metals from group 11 of the periodic table. 
This indicates a surface-related and incoherent origin of CL at 1\,keV. 
The CL spectra of Au and Cu exhibit well-resolved peaks around 500\,nm and 590\,nm, respectively, superimposed on the broadband transition radiation spectrum. 
These results align well with previous CL measurements on thin metal films~\cite{Borziak1976}, suggesting that this radiation stems from interband recombination in Au and surface-related emission of natural oxide layers in Ag and Cu~\cite{Papanicolaou1976}.
The post-transition metals Al and Ti also display isotropic emission profiles at 1\,keV, likely originating from their natural oxide surface layers. 
Finally, the transition metals from the group 10 of periodic table Pt, Ni, and Cr, exhibit the broad transition radiation spectra at 1\,keV (Fig.~\ref{fig-spec-emetals}\pnl{f-h}) with the characteristic doughnut-shaped angular radiation profile (blue in Fig.~\ref{fig-spec-emetals}\pnl{n-p}). 
This is likely because they have closed orbitals, making them chemically inert under normal conditions and preventing the formation of oxides or other surface compounds.

\begin{figure}[t]
\includegraphics[width=0.99\linewidth]{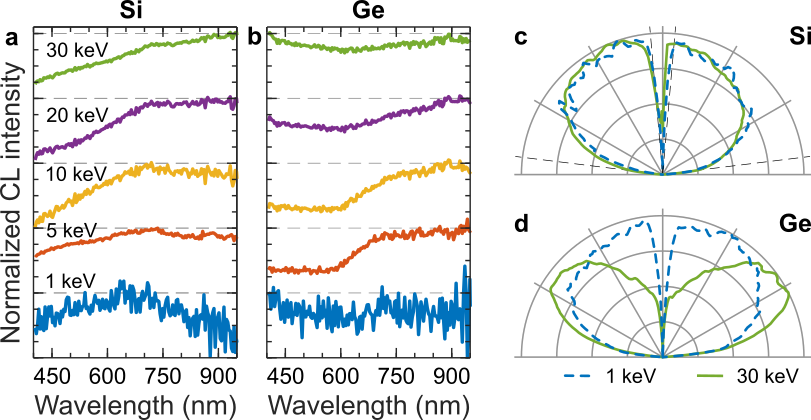}
\caption{
    \textbf{CL of elemental metalloids.} 
    Normalized CL spectra measured in a range of electron beam energies for:
    \textbf{a}~Si, and  
    \textbf{b}~Ge.
    \textbf{c--d}~Corresponding angular radiation profiles measured at 1\,keV~(dashed blue) and 30\,keV~(solid green).
}
\label{fig-spec-si-ge}
\end{figure}

\subsection{5.2 Elemental metalloids}

Elemental metalloids silicon (Si) and germanium (Ge) are group 14 elements in the periodic table, sharing the same valence electron configuration and exhibiting semiconductor behavior with an indirect bandgap.
Previous CL studies of Si and Ge focused on their optical and electronic properties, especially concerning defect states, impurity levels, strain distributions, offering a high spatial resolution method to assess carrier lifetimes and electronic transitions~\cite{Myhajlenko1983,Yacobi1986}. 
These experiments used significantly higher beam currents ranging from 100\,nA to 10\,$\upmu$A, with electron energy adjustments allowing for depth profiling. 

We report CL from bulk Si and Ge crystals in Fig.~\ref{fig-spec-si-ge}\pnl{a,b}, generated at low beam current of ca.~1.4\,nA, in contrast to the previous studies. 
Fig.~\ref{fig-spec-si-ge}\pnl{c,d} show angular radiation profiles of Si and Ge, which are a linear mix of isotropic and dipolar emission patterns~\cite{Brenny2014,Mignuzzi2018}.
The isotropic contribution from incoherent CL originates from indirect bandgap transitions in these materials with bandgap energies corresponding to the IR part of electromagnetic spectrum. 
The dipole emission from coherent CL stems from the transition radiation, which dominates at shorter wavelengths. 
We confirm this in spectral filtering experiments by separating the incoherent and coherent CL with optical filters~{(SM Fig.~S11)}. 
In summary, the CL emission of bulk Si and Ge consists of a combination of incoherent isotropic radiation from indirect bandgap transitions and coherent dipole-like emission from transition radiation, which can be separated using spectral filtering.

\begin{figure}[t!]
\includegraphics[width=1\linewidth]{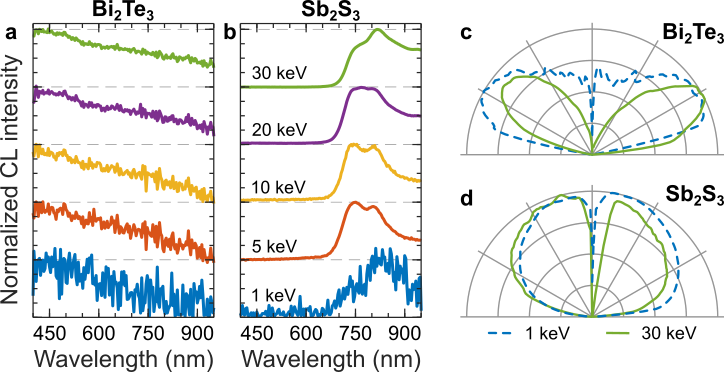}
\caption{\textbf{CL of topological insulators.}	
Normalized CL spectra measured in a range of electron beam energies for:
\textbf{a}~Bi$_2$Te$_3$, and
\textbf{b}~Sb$_2$S$_3$.
\textbf{c--d}~Corresponding angular radiation profiles measured at 1\,keV~(dashed blue) and 30\,keV~(solid green). 
}
\label{fig-spec-topins}
\end{figure}

\begin{figure*}[t]
\includegraphics[width=0.99\linewidth]{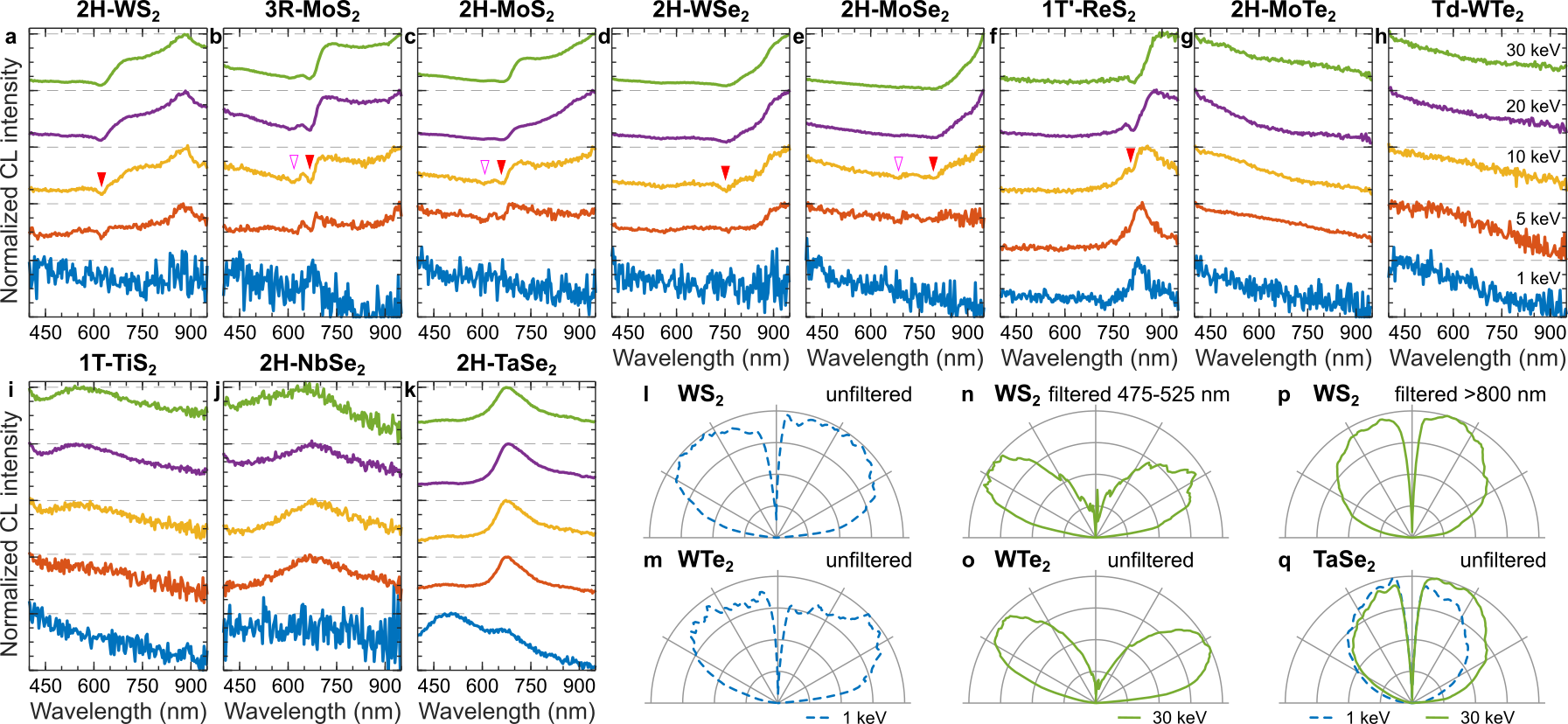}
\caption{\textbf{CL of TMD crystals.}	
Normalized CL spectra measured in a range of electron beam energies 
for a variety of TMD crystals: 
\textbf{a}~2H-WS$_2$, 
\textbf{b}~3R-MoS$_2$,
\textbf{c}~2H-MoS$_2$,
\textbf{d}~2H-WSe$_2$,
\textbf{e}~2H-MoSe$_2$,
\textbf{f}~1T'-ReS$_2$,
\textbf{g}~2H-MoTe$_2$,
\textbf{h}~Td-WTe$_2$,
\textbf{i}~1T-TiS$_2$,
\textbf{j}~2H-NbSe$_2$, and
\textbf{k}~2H-TaSe$_2$. 
Spectral dips corresponding to excitons X$_\mathrm{A}$ (red triangle) and X$_\mathrm{B}$ (open magenta triangle) are indicated for relevant TMD crystals.
\textbf{l--m}~Mixed angular radiation profiles of WS$_2$ and WTe$_2$ at 1\,keV.
\textbf{n,p}~Spectrally decomposed angular radiation profiles of WS$_2$ at 30\,keV into \textbf{n} coherent and \textbf{p} incoherent CL, which are representative for the semiconducting TMDs.
\textbf{o}~Dipolar angular radiation profile of WTe$_2$, which is representative for the semimetallic TMDs.
\textbf{q}~Isotropic angular radiation profile of TaSe$_2$ at 1\,keV~(dashed blue) and 30\,keV (solid green).
}
\label{fig-spec-tmds}
\end{figure*}

\subsection{5.3 Topological insulators}
Topological insulators are materials that behave as insulators in their bulk form but have conductive states on their surfaces, protected by topological order, making them suitable for applications in quantum computing and photonics~\cite{Qi2011}.
CL investigation of topological insulators allows for probing of their electronic band structure properties, defect states, exciton polaritons, and surface phenomena, providing insights into the unique optical and quantum properties of topological states~\cite{ou2014,Lingstdt2021,Yan2024,Mediavilla-Martinez2024}.

We report CL response of bismuth telluride (Bi$_2$Te$_3$) and antimony trisulfide (Sb$_2$S$_3$) crystals in
Fig.~\ref{fig-spec-topins}\pnl{a,b}. 
Bi$_2$Te$_3$ in Fig.~\ref{fig-spec-topins}\pnl{a} exhibits broadband CL similar to elemental metals indicating on transition radiation origin of CL signal, which is further confirmed by the characteristic doughnut-shaped angular radiation profile in Fig.~\ref{fig-spec-topins}\pnl{c}. 
In contrast, Sb$_2$S$_3$ crystal exhibits bright CL with spectrum peaking in the near IR spectral region~(Fig.~\ref{fig-spec-topins}\pnl{b}), corresponding to the quasi-direct bandgap emission~\cite{Zhong2017}. 
The CL of Sb$_2$S$_3$ is characterized by isotropic angular radiation pattern (Fig.~\ref{fig-spec-topins}\pnl{d}), confirming the incoherent, bandgap-related origin.

\begin{figure*}[t]
\includegraphics[width=1\linewidth]{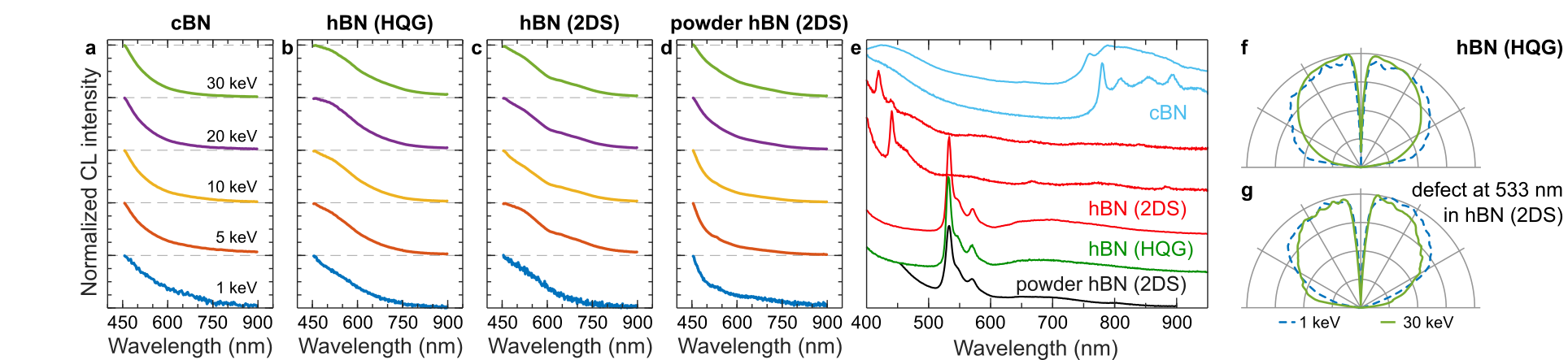}
\caption{\textbf{CL of cBN and hBN crystals.}	
Normalized CL spectra measured in a range of electron beam energies for:
\textbf{a}~micro-crystals of cBN;
\textbf{b}~hBN crystal from HQ Graphene (HQG);
\textbf{c}~hBN crystal from 2D Semiconductors (2DS);
\textbf{d}~hBN power from 2DS.
\textbf{e}~Normalized CL spectra of various defects and color centers measured from cBN (blue), hBN from 2DS (red), hBN from HQG (green), and hBN powder from 2DS (black).
\textbf{f--g}~Isotropic angular radiation profiles measured at 1\,keV~(dashed blue line) and 30\,keV~(solid green line) for hBN crystal from HQG and electron-beam-induced defect emission in hBN crystal from 2DS.
}
\label{fig-spec-BN}
\end{figure*}

\subsection{5.4 TMD crystals}\label{TMD Section}
Transition metal dichalcogenide (TMD) materials are 2D layered crystals, where the layers are weakly bonded by van der Waals forces, allowing them to be exfoliated down to a monolayer. 
Electron beam excitation is inherently inefficient in TMD monolayers due to their extreme thinness, which results in an exceptionally short electron interaction length.
Encapsulation with hexagonal boron nitride (hBN) increases the interaction length, enabling electron-hole pairs to form in the hBN, which then decay through the TMD monolayer generating CL~\cite{Zheng2017}.
Such CL studies have focused on exciton dynamics in monolayers, strain effects, and electron-hole interactions at the nanoscale, providing insights into their optical properties~\cite{Zheng2017,Nayak2019,Bonnet2021,Francaviglia2022,Fiedler2023tmd,Borghi2024,Sun2024}. 
CL methods have been also applied in nano-optics research involving multilayer TMD flakes, which support various optical modes depending on their shape and thickness~\cite{Taleb2021,Chahshouri2022,Vu2022,Vu2023,Woo2024}. 
However, the CL response of bulk TMD crystals remains relatively underexplored, yet it is essential for establishing a reliable reference for CL experiments in photonics and plasmonics.

We report the CL emission properties of bulk TMDs in Fig.~\ref{fig-spec-tmds}, including semiconductor TMDs such as tungsten disulfide (WS$_2$), molybdenum disulfide (MoS$_2$) [both the rhombohedral (3R) and the hexagonal (2H) symmetry phases], tungsten diselenide (WSe$_2$), molybdenum diselenide (MoSe$_2$), rhenium disulfide (ReS$_2$), as well as semimetal and metal TMDs such as molybdenum ditelluride (MoTe$_2$), tungsten ditelluride (WTe$_2$), titanium disulfide (TiS$_2$), niobium diselenide (NbSe$_2$), and tantalum diselenide (TaSe$_2$). 
For each investigated TMD crystal we identified its symmetry phase based on Raman spectra~{(see SM).} 

CL response of semiconducting TMD crystals (WS$_2$, MoS$_2$, WSe$_2$, MoSe$_2$, ReS$_2$) exhibits a complex structure~(Fig.~\ref{fig-spec-tmds}\pnl{a-f}), combining coherent and incoherent contributions as revealed from angular spectra measurements.
Angular radiation profiles both at 1\,keV and 30\,keV measured in the broadband spectral range exhibit mixed isotropic-dipole patterns~as in Fig.~\ref{fig-spec-tmds}\pnl{l} for WS$_2$, while the spectral filtering experiments help to decompose them in coherent and incoherent contributions as in Fig.~\ref{fig-spec-tmds}\pnl{n,p}.
CL intensity dominates at wavelengths longer than the expected A exciton line X$_\mathrm{A}$ in Fig.~\ref{fig-spec-tmds}\pnl{a-f}~(marked by the red filled triangle), and we attribute this emission to electron-hole recombination near the bandgap energy.
Therefore, this bandgap-related emission is characterized by incoherent CL with isotropic angular radiation profile, which we demonstrate on the example of a WS$_2$ crystal using a long-pass filter at 800\,nm in Fig.~\ref{fig-spec-tmds}\pnl{p}.
In contrast, at the wavelength below X$_\mathrm{A}$, CL emission exhibits perfectly dipolar angular radiation profile, as shown for WS$_2$ crystal measured with a band-pass filter at 475--525\,nm in Fig.~\ref{fig-spec-tmds}\pnl{n}.
We repeated these optical filtering experiments for all the semiconducting TMD crystals from Fig.~\ref{fig-spec-tmds}\pnl{a-f}, confirming these CL emission properties {(SM Fig.~S12).}

Interestingly, we observe spectral dips at the position of A and B excitons in the semiconducting TMD crystals~(indicated by the filled red and open magenta triangles in Fig.~\ref{fig-spec-tmds}\pnl{a-f}, respectively).
We attribute these spectral features to self-absorption of bandgap-related emission in bulk TMD flakes by A and B excitons, which are consistent with refractometry measurements~\cite{Munkhbat2022}~{(SM Fig.~S19)}. 
This is further supported by CL experiments of few layer MoS$_2$ on emissive substrates, where spectral dips were observed around the exciton energies~\cite{Negri2022}.
Additionally, similar spectral dips have been reported in CL studies of thin TMD crystal slabs, attributed to strong exciton-photon coupling between cavity modes and excitons~\cite{Taleb2021,Chahshouri2022}. 

Semimetallic and metallic TMD crystals (MoTe$_2$, WTe$_2$, TiS$_2$, NbSe$_2$) exhibit broadband CL emission (Fig.~\ref{fig-spec-tmds}\pnl{g-j}), similar to the CL response of metals.
This CL emission has the coherent origin as confirmed by the dipolar angular radiation profiles as shown on the example of WTe$_2$ in Fig.~\ref{fig-spec-tmds}\pnl{o}~{(see also SM Fig.~S12)}.
In contrast, the TaSe$_2$ crystal is a metallic TMD that exhibits strong photoluminescence due to interband transitions~\cite{Mahajan2019}.
We observe a similar response of TaSe$_2$ in CL with strong emission around~675\,nm~and isotropic angular radiation profile~(Fig.~\ref{fig-spec-tmds}\pnl{k,q}).
\begin{figure*}[t!]
\includegraphics[width=1\linewidth]{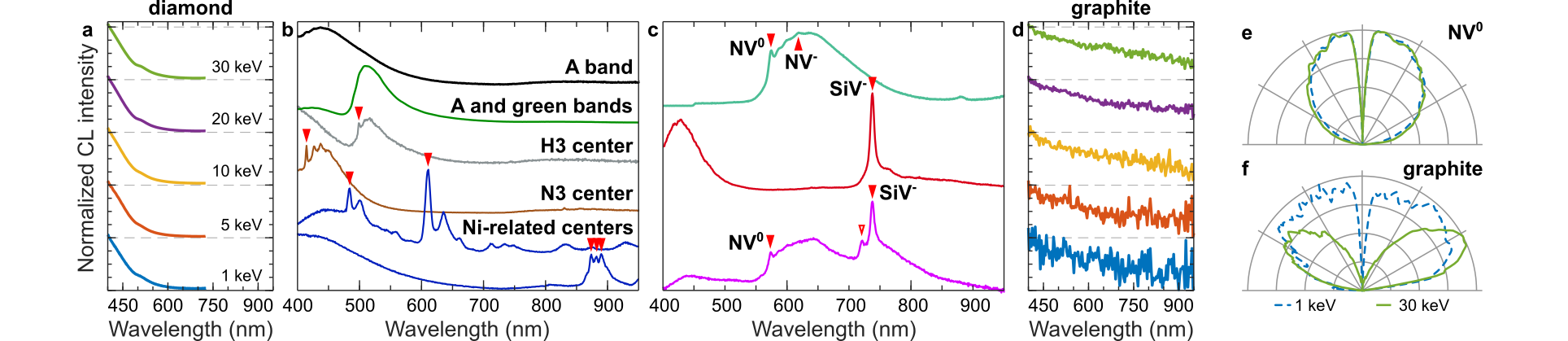}
\caption{\textbf{CL of diamond and graphite.}	
\textbf{a}~Normalized CL spectra of a diamond crystal (ThermoFisher), measured at various electron beam energies. Data above 725\,nm was excluded due to strong second-order diffraction artifacts.
\textbf{b}~Defect emission in diamond crystals (30--60\,$\upmu$m across, ThermoFisher). 
\textbf{c}~Color centers in doped nanodiamonds (<1\,$\upmu$m across, Adamas Technologies).
The spectral positions of the emission lines discussed in the main text are marked with red triangles.
\textbf{d}~Normalized CL spectra of graphite crystal, measured at various electron beam energies (see panel \textbf{a} for color coding).
\textbf{e}~Isotropic angular radiation profile of NV$^0$ color center in a diamond nanocrystal at 1\,keV~(dashed blue) and 30\,keV~(solid green), respectively, measured with a short pass filter at 600\,nm.
\textbf{f}~Mixed angular radiation pattern at 1\,keV~(dashed blue) and dipolar at 30\,keV~(solid green) of graphite.
}
\label{fig-spec-carbon}
\end{figure*}

\subsection{5.5 Boron nitride}

Boron nitride (BN) exists in various structural forms, with cubic (cBN) and hexagonal (hBN) being the most commonly used in photonics due to their distinct optical and electronic properties. 
cBN has a structure similar to diamond, while hBN is a van der Waals layered material analogous to graphite.
Both cBN and hBN have a wide bandgap (5.9--6.4\,eV), which can host a variety of narrow-line emitters operating at room temperature, making the BN platform highly suitable for quantum emitter engineering~\cite{Shipilo1986,Silly2007,Bourellier2016,tararan2018,Shevitski2019,Hayee2020,Roman_2021,Gale2022,Roux2022,Bianco2023}.
The high spatial resolution of CL methods aids in understanding defect-related emission while also providing insights into exciton behavior and phonon interactions at the nanoscale~\cite{Taylor1994,Zhang2002,Watanabe2004,Schu2016,Bogroff2025}.
Additionally, the electron beam can create or activate defects in hBN~\cite{Gale2022,Roux2022,Bianco2023,Bogroff2025}, unlocking potential for deterministic quantum emitter generation and integration with photonic and quantum devices.
Furthermore, encapsulating TMD monolayers with hBN has enabled their efficient electron beam excitation~\cite{Zheng2017}. 

We report the CL response of cBN and various hBN crystals sourced from different suppliers, along with observed narrow-line emission defects~(Fig.~\ref{fig-spec-BN}).
Fig.~\ref{fig-spec-BN}\pnl{a-d} present CL spectra of cBN and hBN crystals, which exhibit strong CL emission in the UV spectral region, in good agreement with previous CL studies~\cite{Taylor1994,Zhang2002,Watanabe2004}.
We observe an isotropic angular radiation profile for this CL signal, as demonstrated on the example of an hBN crystal from HQ Graphene in Fig.~\ref{fig-spec-BN}\pnl{f}~{(see also SM Fig.~S13)}.

We examined multiple cBN and hBN crystals to identify defect sites characterized by narrow-line emission, as summarized in~Fig.~\ref{fig-spec-BN}\pnl{e}.
In cBN micro-crystals (light blue in Fig.~\ref{fig-spec-BN}\pnl{e}), we observe defect-related emission in the near-IR spectral range around 759\,nm (1.634\,eV), which may be associated with defects reported in Refs.~\cite{Shipilo1986,tararan2018}, along with a more prominent defect emission centered around 770\,nm~(1.610\,eV), which, to the best of our knowledge, has not been previously reported.
In hBN crystals from 2D~Semiconductors~(red in Fig.~\ref{fig-spec-BN}\pnl{e}), we observe blue defects with a CL spectrum centered around 419\,nm and 440\,nm, which are related to nitrogen interstitial defects or B centers and can be electron-beam induced~\cite{Roux2022,Ganyecz2024}.
We observe similar defects in hBN crystals from HQ Graphene~(green in Fig.~\ref{fig-spec-BN}\pnl{e}) and in hBN micropowder from 2D~Semiconductors~(black in Fig.~\ref{fig-spec-BN}\pnl{e}).
Finally, we observe a bright defect emission centered around 533\,nm in hBN crystals from all suppliers, exhibiting an isotropic angular radiation profile~(Fig.~\ref{fig-spec-BN}\pnl{g}). 
Activation of this defect by an electron beam has been reported in Ref.~\cite{Bianco2023}, whereas we were able to induce its formation using an electron beam at ca.~1.4\,nA in mechanically exfoliated hBN crystals regardless of the sample source (HQ~Graphene or 2D~Semiconductors).

\subsection{5.6 Diamond and graphite}

Carbon forms a variety of allotropes, such as diamond and graphite, each exhibiting unique and rich optoelectronic properties~\cite{Zaitsev2001,Falcao2007}.
Diamond, an insulator with a wide bandgap (5.5\,eV), is among the most extensively studied materials in CL microscopy~\cite{Woods1975,Yacobi1990}.
It serves as a prominent platform for quantum emitter engineering, as its wide bandgap can host stable narrow-line emitters at room temperature, making it valuable for quantum technology applications~\cite{Bradac2019}. 
Additionally, CL from these defects enables the study of the fundamental physical properties of electron-induced incoherent light emission~\cite{Tizei2013,Meuret2015,Fiedler2023,Iyer2023}. 
Graphite, the hexagonal allotrope of carbon, exhibits semimetallic characteristics and has primarily been studied for its CL properties in geological investigations~\cite{Kostova2012,GarciaGuinea2018}.

We report the CL response of synthetic diamond crystals from ThermoFisher and Adamas Nanotechnologies~(Fig.~\ref{fig-spec-carbon}). 
The ThermoFisher diamonds are relatively large, measuring tens of microns across, and are not intentionally doped, allowing us to study the bulk CL response of diamond. 
Their CL spectra, shown in Fig.~\ref{fig-spec-carbon}\pnl{a}, peak in the UV region beyond our detection range.
The bulk CL response is often accompanied by broad defect emission around 430\,nm and 515\,nm~(black and green in Fig.~\ref{fig-spec-carbon}\pnl{b}), which correspond to the A and green bands, respectively~\cite{Zaitsev2001}.
Occasionally, these crystals also exhibit narrow-line defect or impurity emission, among which we observe emission from H3, N3, and Ni-related centers~(marked by red triangles in Fig.~\ref{fig-spec-carbon}\pnl{b}).

The diamonds from Adamas Nanotechnologies are submicron in size and are intentionally doped with NV and SiV centers at varying densities, enabling the isolation of single-photon emitters. 
Fig.~\ref{fig-spec-carbon}\pnl{c} presents the CL spectra of NV$^0$ (574\,nm), NV$^-$ (638\,nm), SiV$^-$~(738\,nm), and Si-related~(722\,nm) centers~\cite{Zaitsev2001,Sedov2016,Becker2020,Fiedler2023}, with some diamonds containing a mixture of these centers. 
Fig.~\ref{fig-spec-carbon}\pnl{e} shows a representative isotropic angular radiation profile of the NV$^0$ center~{(see also SM Fig.~S14)}.

In contrast to diamond, the CL of graphite is much dimmer and is the weakest among all investigated materials in our atlas.
Fig.~\ref{fig-spec-carbon}\pnl{d} shows the broadband CL spectrum of a graphite crystal, resembling the transition radiation observed in metals.
The corresponding angular radiation profiles in Fig.~\ref{fig-spec-carbon}\pnl{f} are similar to those observed in metals and metalloids. 
The angular radiation profile at 30\,keV is dipolar (green in Fig.~\ref{fig-spec-carbon}\pnl{f}), and, together with the broadband spectrum, suggests a coherent CL origin attributed to transition radiation. 

\begin{figure*}
\includegraphics[width=0.99\linewidth]{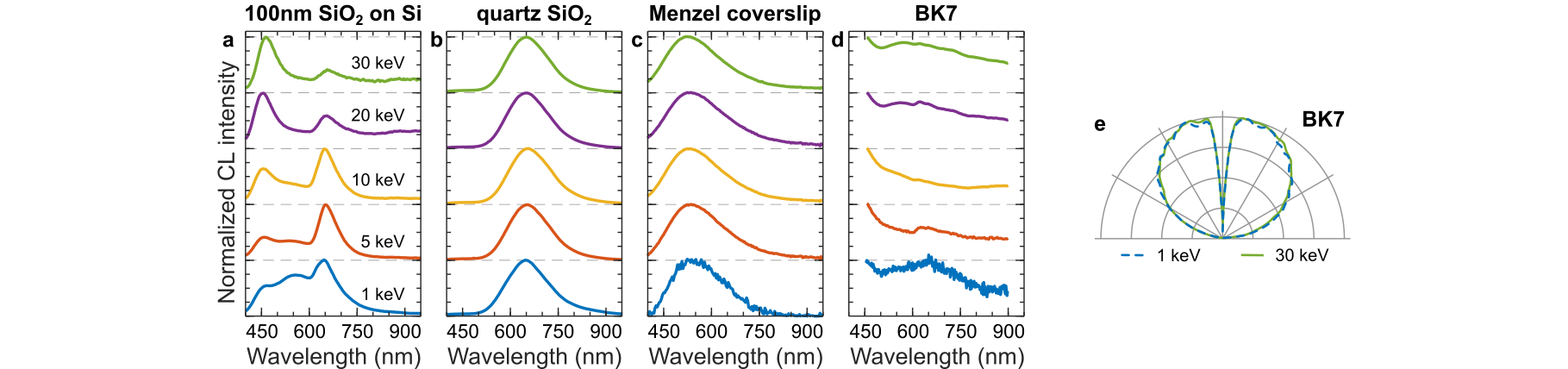}
\caption{\textbf{CL of common glasses.} 
    Normalized CL spectra measured in a range of electron beam energies for:
    \textbf{a}~100\,nm-thick layer of SiO$_2$ on bulk Si, 
    \textbf{b}~$\alpha$-quartz (SiO$_2$), 
    \textbf{c}~Menzel coverslip, and 
    \textbf{d}~BK7 glass. 
    \textbf{e}~Isotropic angular radiation profiles of BK7 glass at 1\,keV~(dashed blue) and 30\,keV~(solid green), representative for all materials in panels~\textbf{a-d}. 
}
\label{fig-spec-glasses}
\end{figure*}

\subsection{5.7 Common glasses}

CL methods are widely used for glass characterization, providing insights into their microstructural and optical properties~\cite{Koyama1980,Yacobi1990}. 
We report the CL response of SiO$_2$-based glasses in Fig.~\ref{fig-spec-glasses}. 
CL emission from SiO$_2$ is complex and originates from the luminescence associated with various defects~\cite{Mitchell1973,Koyama1977,Koyama1980,Skuja1986,Kalceff1995}.
Fig.~\ref{fig-spec-glasses}\pnl{a} presents CL spectra of a thin, 100\,nm dry-grown oxide layer on Si, which is commonly used to exfoliate 2D materials, such as graphene and TMD monolayers. 
The three luminescent peaks in Fig.~\ref{fig-spec-glasses}\pnl{a} correspond to emission from non-bridging oxygen hole centers around 460\,nm~\cite{Mitchell1973,FITTING20052251}, self-trapped excitons around 560\,nm~\cite{FITTING20052251}, and oxygen vacancies around 650\,nm~\cite{Mitchell1973,FITTING20052251}.
Fig.~\ref{fig-spec-glasses}\pnl{b} shows CL spectra of $\alpha$-quartz with a dominant peak around 650\,nm, attributed to oxygen vacancies~\cite{Koyama1980,Kalceff1995}. 

Amorphous SiO$_2$ is frequently modified through the incorporation of additional oxides to tailor its optical properties. 
We present CL spectra of SiO$_2$-based borosilicate glasses, specifically the commercial brands Menzel-Gl\"aser and BK7, as shown in Fig.~\ref{fig-spec-glasses}\pnl{c,d}. 
Assigning these CL spectra to specific defect emissions is challenging due to the complex structural composition of borosilicates, but the observed spectral peaks are generally associated with SiO$_2$-related defects~\cite{Stevens-Kalceff_2008}. 
Finally, Fig.~\ref{fig-spec-glasses}\pnl{e} shows isotropic angular radiation profiles of BK7, which are representative of all the glasses presented~{(see also SM Fig.~S15).}

\begin{figure*}
\includegraphics[width=0.99\linewidth]{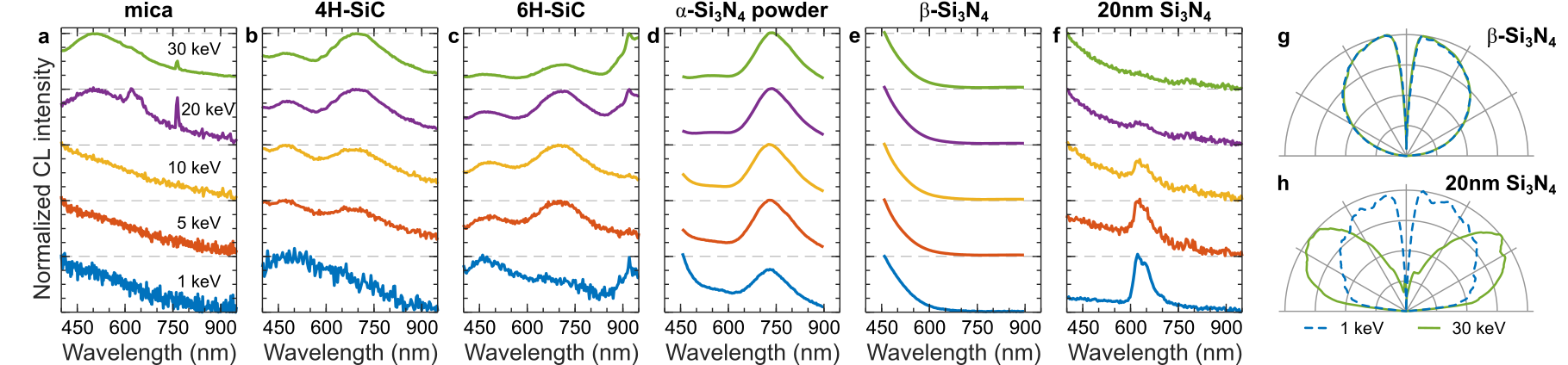}
\caption{\textbf{CL of silicon-based materials.} 
    Normalized CL spectra measured in a range of electron beam energies for:
    \textbf{a}~mica,
    \textbf{b}~4H-SiC, 
    \textbf{c}~6H-SiC,
    \textbf{d}~$\alpha$-Si$_3$N$_4$ nanopowder,
    \textbf{e}~crystalline $\beta$-Si$_3$N$_4$, and   
    \textbf{f}~20\,nm-thin Si$_3$N$_4$ membrane.
    \textbf{g}~Isotropic angular radiation profiles of $\beta$-Si$_3$N$_4$ at 1\,keV~(dashed blue) and 30\,keV~(solid green), representative for materials in panels~\textbf{a-e}.
    \textbf{h}~Isotropic and dipolar angular radiation profiles of 20\,nm-thin Si$_3$N$_4$ membrane at 1\,keV~(dashed blue) and 30\,keV~(solid green), respectively.
}
\label{fig-spec-simaterials}
\end{figure*}

\subsection{5.8 Silicon-based materials}

We report the CL response of silicon-based materials, such as mica, silicon carbide (SiC), and silicon nitride (Si$_3$N$_4$), in Fig.~\ref{fig-spec-simaterials}. 
Mica, a naturally occurring layered silicate, is used as an atomically flat substrate with a wide range of applications in photonics~\cite{Fali2020}, and has also been employed to study Cherenkov and transition radiation generated in a transmission electron microscope~\cite{Yamamoto1996}.
Here we report CL of Muscovite mica (KO$_2$Al$_2$O$_3$SiO$_2$) in Fig.~\ref{fig-spec-simaterials}\pnl{a}, which has broad spectra spanning the visible to near-IR regions with an isotropic angular radiation profile~(SM Fig.~S16).
This is indicative of an incoherent defect-related emission from oxide components in mica; notably, at higher electron beam energies, we observe an additional narrow-line emission around 764~nm.

SiC, a wide-bandgap semiconductor, finds applications in optoelectronics due to its high thermal conductivity and stability. 
SiC can also host a range of CL-active defect centers~\cite{Sorman2000,Kakanakova2002,Ottaviani_2004}, such as the silicon vacancy and divacancy complexes, with potential for quantum information and sensing technologies~\cite{Castelletto2022}. 
While SiC exists in various structural polytypes, here we focus on the 4H and 6H phases, which are widely used in photonics. 
Fig.~\ref{fig-spec-simaterials}\pnl{b,c} show the CL spectra of the 4H and 6H phases, respectively.
These SiC phases exhibit similar spectral features, including two broad emission bands centered around 470\,nm and 680\,nm.
The emission band around 470\,nm originates from boron impurities in SiC, while the band near 680\,nm is associated with basal plane dislocations~\cite{Kakanakova2002,Ottaviani_2004}. 
Additionally, an emission line appears around 918\,nm in the 6H-SiC sample, likely originating from silicon vacancies~\cite{Sorman2000}.

Fig.~\ref{fig-spec-simaterials}\pnl{d-f} presents the CL spectra of Si$_3$N$_4$ in various forms: $\alpha$-Si$_3$N$_4$ micro-powder, compressed crystals of $\beta$-Si$_3$N$_4$, and a 20\,nm-thin Si$_3$N$_4$ membrane. 
$\alpha$- and $\beta$-Si$_3$N$_4$ are used in low-loss photonics and support nano-structuring, having a wide bandgap (4.8--5.2\,eV) and exhibiting defect-related emission in CL~\cite{Hu2004,Sekiguchi2004,Huang2013,Zhu2015}.
Fig.~\ref{fig-spec-simaterials}\pnl{d} presents CL spectra of $\alpha$-Si$_3$N$_4$, with a strong emission band around 730~nm, and a minor peak around 540\,nm, which are likely related to defect emission. 
Our $\alpha$-Si$_3$N$_4$ sample also exhibits intense photoluminescence, preventing phase characterization via Raman microscopy~{(see corresponding spectrum in SM)}.
In contrast, $\beta$-Si$_3$N$_4$ exhibits CL only in the UV region (Fig.~\ref{fig-spec-simaterials}\pnl{e}), and shows no photoluminescence in the visible region.
As a membrane, Si$_3$N$_4$ is commonly used in electron beam spectroscopy as a free-standing platform.
With thicknesses down to 5\,nm, it enables substrate-free CL measurements, minimizing background signals and improving sensitivity for optical mode characterization~\cite{Fiedler2022dis}.
Therefore, its characterization is of particular importance for CL microscopy.
Here we report CL of a 20\,nm-thin Si$_3$N$_4$ membrane in Fig.~\ref{fig-spec-simaterials}\pnl{f}. 
At low electron beam energies, it exhibits a CL peak around 620~nm, attributed to defect-related emission; this peak disappears at higher energies, as electrons pass through the membrane without  interaction.
In contrast, the high energy electrons cause transition radiation at the vacuum-membrane interface~\cite{Yamamoto1996}, which we confirm with angular emission measurements. 
Fig.~\ref{fig-spec-simaterials}\pnl{h} shows an isotropic angular emission profile for defect emission around 620~nm efficiently excited at 1~keV, whereas at 30~keV, a dipolar angular emission profile characteristic of transition radiation is observed.

\subsection{5.9 Common photonic materials and optical coatings}

\begin{figure*}[t]
\includegraphics[width=1\linewidth]{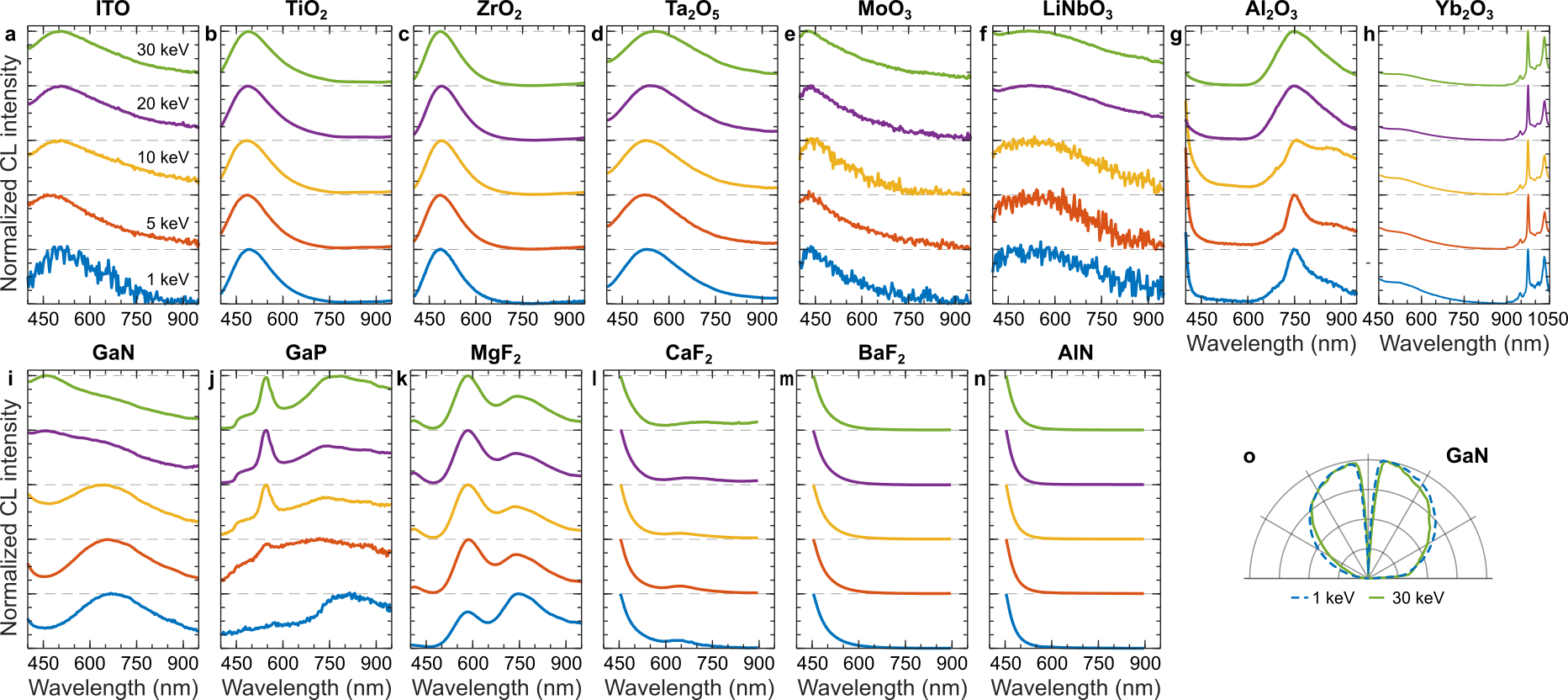}
\caption{\textbf{CL of common photonic materials and optical coatings.}	
Normalized CL spectra measured in a range of electron beam energies for:
\textbf{a}~ITO, 
\textbf{b}~TiO$_2$, 
\textbf{c}~ZrO$_2$, 
\textbf{d}~Ta$_2$O$_5$, 
\textbf{e}~MoO$_3$, 
\textbf{f}~LiNbO$_3$, 
\textbf{g}~$\alpha$-Al$_2$O$_3$, 
\textbf{h}~Yb$_2$O$_3$, 
\textbf{i}~GaN, 
\textbf{j}~GaP, 
\textbf{k}~MgF$_2$, 
\textbf{l}~CaF$_2$, 
\textbf{m}~BaF$_2$, and 
\textbf{n}~AlN.
\textbf{o}~Isotropic angular radiation profiles of GaN at 1\,keV~(dashed blue line) and 30\,keV~(solid green line), representative for all materials in panels~\textbf{a--n}. 
}
\label{fig-spec-others}
\end{figure*}

Understanding bulk CL emission is crucial for differentiating it from geometry-tailored emission arising from optical modes and resonances or from broadband transition radiation.
We report CL response for common photonic materials and optical coatings in Fig.~\ref{fig-spec-others}, covering a broad range of semiconductors and dielectrics employed in photonic and plasmonic devices.
We further divide this section into subsections, each focused on a specific material category, including oxides, gallium derivatives, fluorides, and nitrides.

\textit{Oxides.}~We selected widely used oxides in photonics and nanotechnology such as indium tin oxide (ITO), titanium dioxide (TiO$_2$), zirconium dioxide (ZrO$_2$), tantalum pentoxide (Ta$_2$O$_5$), molybdenum trioxide (MoO$_3$), lithium niobate (LiNbO$_3$), aluminum oxide (Al$_2$O$_3$), and ytterbium oxide (Yb$_2$O$_3$), presenting their respective CL spectra along with their interpretation.

Fig.~\ref{fig-spec-others}\pnl{a} shows CL spectra of ITO, which combines optical transparency and electrical conductivity, functions as an epsilon-near-zero (ENZ) material, and offers tunable plasmonic properties in the near IR. 
The broad CL spectra from ITO are characterized by an emission band centered around 500\,nm, attributed to its indirect bandgap and varies between 2--3.5\,eV, depending on the exact doping composition of ITO~\cite{Maestre2008,PAN2013}.

Fig.~\ref{fig-spec-others}\pnl{b} shows CL spectra of anatase-TiO$_2$, which exhibits a high refractive index, wide bandgap, strong photocatalytic activity, and the ability to support localized surface plasmon resonances when doped.
The CL from anatase exhibits an emission band centered around 490\,nm, likely related to emission from Ti$^{3+}$ ions, while the tail in the visible region is associated with oxygen vacancy centers and oxygen interstitials~\cite{Battiston2010,Santara2013,Machreki2023}.

Fig.~\ref{fig-spec-others}\pnl{c} shows CL spectra of ZrO$_2$, a high-temperature, chemically stable ceramic with a high dielectric constant and optical transparency.
The CL from ZrO$_2$ exhibits an emission band centered around 485\,nm, originating from intrinsic oxygen vacancies within the bandgap~\cite{Bofelli2014,Hur2012}.

Fig.~\ref{fig-spec-others}\pnl{d} shows CL spectra of Ta$_2$O$_5$, a high-refractive-index dielectric with excellent insulating properties and transparency from the visible to IR range.
The CL from Ta$_2$O$_5$ exhibits a broad emission band centered around 560\,nm at 30\,keV, slightly blueshifting to 540\,nm at 1\,keV, likely originating from defects within the oxide structure, such as oxygen vacancies or tantalum luminescence centers~\cite{Baraban2016}.
As with other oxides, oxygen vacancy formation within the 3.75\,eV bandgap can strongly influence its optical response~\cite{Lee2017}, contributing to the observed CL.

Fig.~\ref{fig-spec-others}\pnl{e} shows CL spectra of MoO$_3$, which is characterized by a high refractive index, tunable bandgap, strong birefringence, and the ability to support surface plasmon polaritons.
The CL from MoO$_3$ exhibits a broad emission band spanning from the UV to the visible region, originating from its 3\,eV bandgap~\cite{Zhao2003,Koike2014}.

Fig.~\ref{fig-spec-others}\pnl{f} shows CL spectra of LiNbO$_3$, which is characterized by a high refractive index, strong nonlinear optical coefficients, and excellent electro-optic properties.
The CL from LiNbO$_3$ exhibits a broad emission band peaking around 540\,nm and a shoulder extending into the near-IR, primarily originating from defect-related emission~\cite{Llopis1984}.

Fig.~\ref{fig-spec-others}\pnl{g} shows CL spectra of $\alpha$-Al$_2$O$_3$, the crystalline polymorph of sapphire. 
It is characterized by high transparency across a wide spectral range, exceptional hardness, high thermal stability, and excellent dielectric properties.
The CL from $\alpha$-Al$_2$O$_3$ features a band in the UV region related to oxygen vacancies center and a band centered around 750\,nm attributed to Ti$^{3+}$ impurities~\cite{Ghamnia1997,demol2019}. 
The band at 750\,nm is broad at 30\,keV and becomes narrower as the electron energy decreases.
We also observe a narrow emission line around 690\,nm, attributed to Cr$^{3+}$ impurities~\cite{Ghamnia1997,demol2019}.

Fig.~\ref{fig-spec-others}\pnl{h} shows CL spectra of Yb$_2$O$_3$, which is characterized by a high refractive index, a wide bandgap, efficient luminescence, and a high rare-earth doping potential~\cite{Dikovska2004}. 
The CL from Yb$_2$O$_3$ exhibits a broad spectral band centered in the UV spectral region.
Additionally, we observe bright narrow spectral lines at 947\,nm, 976\,nm, 1009\,nm, and 1033\,nm in the near-IR region; due to their high brightness, the detection range for Yb$_2$O$_3$ was extended up to 1050\,nm in Fig.~\ref{fig-spec-others}\pnl{h}.
These lines align well with the emission spectrum of the 4f-electron levels of Yb$^{3+}$, where the brightest emission at 976\,nm corresponds to the ${^2F}_{7/2}\rightarrow {^2F}_{5/2}$ transition~\cite{Buchanan1967}.

\textit{Gallium derivatives.}~Gallium (Ga) is primarily used in the form of III-V compound semiconductors, such as the gallium nitride (GaN) and gallium phosphide (GaP).
Fig.~\ref{fig-spec-others}\pnl{i} shows CL spectra of GaN, which is characterized by a wide direct bandgap, high thermal conductivity, strong emission in the UV-visible range, and the ability to support surface plasmon polaritons.
The CL exhibits a broad emission band centered around 630\,nm at 1\,keV relating to the ubiquitous defect and
impurity emission in GaN~\cite{Mattila1996,Kucheyev2001,Lei2002}.

Fig.~\ref{fig-spec-others}\pnl{j} shows CL spectra of GaP, which is characterized by an indirect bandgap with a high refractive index, high optical transparency, and strong dielectric properties, making it attractive material for optical nanoantennas.
The CL from GaP exhibits a narrow peak at 548\,nm superimposed on a broad band centered around 750--800\,nm.
The emission peak at 548\,nm is characteristic of free exciton recombination from the indirect bandgap of GaP at room temperature~\cite{Wight1977}. 
The broadband emission can originate from various impurities and defects~\cite{Dimitriadis1978,StevensKalceff2001}. 
Depending on the electron beam energy in Fig.~\ref{fig-spec-others}\pnl{j}, a different composition of impurities and defects alters the emission spectrum.

\begin{figure*}[ht!]
\includegraphics[width=1\linewidth]{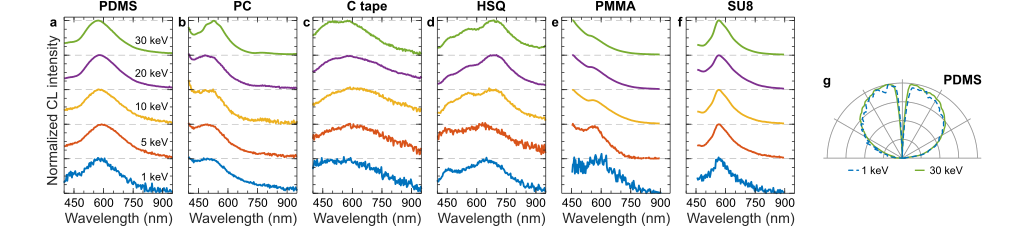}
\caption{\textbf{CL of polymeric and resist materials.}	
Normalized CL spectra measured in a range of electron beam energies 
for:
\textbf{a}~PDMS,
\textbf{b}~PC,
\textbf{c}~carbon tape,
\textbf{d}~HSQ,
\textbf{e}~PMMA, and
\textbf{f}~SU8.
\textbf{g}~Isotropic angular radiation profiles of PDMS at 1\,keV~(dashed blue) and 30\,keV~(solid green), representative for all materials in panels~\textbf{a--f}.
}
\label{fig-spec-polymers}
\end{figure*}

\textit{Fluorides.}~Fluoride materials have emerged as an important class of non-oxide glasses, offering a wide optical transmission bandwidth from the UV to IR, a low refractive index, and excellent durability, enabling numerous applications in photonics. 
We report CL of magnesium fluoride (MgF$_2$), calcium fluoride (CaF$_2$), and barium fluoride (BaF$_2$) in Fig.~\ref{fig-spec-others}\pnl{k-m}.
The observed CL emission originates from structural defects and vacancy centers~\cite{Topaksu2016,Rodrguez2021,HisashiYoshida2000}.

\textit{Nitrides.}~We have previously presented nitride materials such as BN, Si$_3$N$_4$, and GaN. 
Here, we conclude with aluminum nitride (AlN), which is widely used in photonics due to its large bandgap of approximately 6.6\,eV and broad transmission range~\cite{Li2021}. 
The cathodoluminescence of AlN is characterized by a bright emission coming from the UV spectral region (see Fig.~\ref{fig-spec-others}\pnl{n}) and is usually attributed to the emission from oxygen color centers and aluminum vacancies~\cite{Youngman1990,Mattila1996,demol2019}.

Finally, Fig.~\ref{fig-spec-others}\pnl{o} shows the isotropic angular radiation profile of GaN, representative of all materials in this section, and indicating the incoherent origin of CL~{(see also SM Fig.~S17).}

\subsection{5.10 Polymeric and
resist materials}

Polymeric and resist materials are widely used in photonics, plasmonics, and nanotechnology fabrication as patternable substrates for lithography, enabling the creation of fine nanoscale features and complex optical structures.
The materials selected here include polydimethylsiloxane (PDMS), poly(bisphenol A carbonate) or polycarbonate (PC), carbon tape (SPI Supplies), hydrogen silsesquioxane (HSQ), poly(methyl methacrylate) (PMMA), and the commercial photoresist SU8-2000 (Kayaku Advanced Materials). 
Awareness of their respective CL signals is of high importance, as these materials may serve as components in photonic and plasmonic devices or appear as residuals within them.
Reliable CL measurements in polymeric materials require a tailored approach to prevent rapid degradation from extended electron beam exposure, which can significantly reduce signal intensity~\cite{Wellmann2007, Bischak2015}. 
An effective approach to mitigate this challenge is scanning a larger surface area of the polymer, capturing luminescence over a wider region to preserve signal quality while minimizing material damage.

The CL in polymers can arise from various mechanisms, one of which involves the formation of excitonic states that, upon relaxation to their ground state, generate light~\cite{Wellmann2007,Qiao2016}. 
The electron beam may also induce impact ionization of molecules, leading to luminescent recombination with secondary and backscattered electrons~\cite{Tyutnev1983,Qiao2016}, or the formation of a new luminescent molecule as a result of a chemical transformation~\cite{Barrios2012}.
These mechanisms exclude coherent CL processes in the emission from polymers that result in completely isotropic emission profiles in the angular resolved emission (see Fig.~\ref{fig-spec-polymers}\pnl{g} and~{SM Fig.~S18}).

Fig.~\ref{fig-spec-polymers} presents CL response of selected polymeric and resist materials. 
Fig.~\ref{fig-spec-polymers}\pnl{a} shows the CL from PC, a thermoplastic polymer containing carbonate groups, commonly used in the pick-up assembly of van der Waals heterostructures~\cite{Wang2013}.
The CL of PC is characterized by a broad peak around 530\,nm (Fig.~\ref{fig-spec-polymers}\pnl{a}).

Fig.~\ref{fig-spec-polymers}\pnl{b} shows the CL of PDMS, a silicon-based polymer with high optical transparency, elasticity, and heat resistance. 
PDMS is widely used for the exfoliation of TMD monolayers, graphene, and 2D material  assembly~\cite{Wang2013,CastellanosGomez2014}, as well as for the fabrication of layered organic optoelectronic devices~\cite{Yang2017}.
PDMS exhibits bright and broad CL, with a peak around 595\,nm and a shoulder centered around 460\,nm.

Carbon tape is a conductive carbon-acrylic mixture used for mounting samples in electron beam-based techniques. 
Fig.~\ref{fig-spec-polymers}\pnl{c} shows the CL signal of carbon tape, which exhibits a broad emission band centered around 590\,nm.

Fig.~\ref{fig-spec-polymers}\pnl{d} shows the CL response of HSQ, a resist material with a cage-like structure and chemical composition of [HSiO${_{3/2}}$]$_{2n}$~\cite{Grigorescu2009}.
The CL spectra of HSQ in Fig.~\ref{fig-spec-polymers}\pnl{d} exhibit features as other SiO$_2$ derivatives in Fig.~\ref{fig-spec-glasses}. 
CL from HSQ has been observed in nanostructures used as a platform for generating sub-keV electron beam-induced Smith-Purcell radiation~\cite{Roitman2024}.

PMMA is a thermoplastic resist material that is used in photo and electron beam lithography~\cite{Chou1996}. 
The CL from PMMA is shown in Fig.~\ref{fig-spec-polymers}\pnl{e}, exhibiting bright emission in the UV spectral region and extending into the visible region. 
We also observe a local emission maximum around 560\,nm, attributed to emissions from C=C bonds formed due to electron irradiation~\cite{Barrios2012}.

Fig.~\ref{fig-spec-polymers}\pnl{f} shows CL spectra of SU8-2000, a photo- and electron-beam resist material~\cite{Aktary2003}. 
The CL from SU8 features a broad emission band centered around 565\,nm.

\section{6 Cathodoluminescence intensity}
Now we turn our discussion to the response of CL intensity to electron beam current and energy.
The dependence of CL intensity on electron beam current is typically linear~\cite{Wittry1967,Yacobi1990,Bok2014}, as higher currents increase the probability of electron interactions with the material, leading to brighter luminescence. 
However, this trend may deviate depending on the CL mechanism, exhibiting sub- as well as super-linear trends~\cite{RaoSahib1969,Kucheyev2001,Bittorf2025}.
The relationship between CL intensity and electron beam energy can follow a more complex trend. 
At low beam energies, CL intensity is primarily influenced by shallow electron penetration and scattering effects. 
In contrast, at higher energies, a larger material volume contributes to radiative recombination, while deeper electron penetration reduces energy loss to backscattered and secondary electrons~\cite{Bok2014}.
Moreover, at higher electron beam energies, photons are generated deeper within the material, increasing their travel distances and the likelihood of reabsorption. 
As a result, CL intensity may saturate or even decrease due to self-absorption effects, particularly in materials with high optical absorption coefficients~\cite{Koch1988}.
Understanding CL intensity dependencies on current and energy is necessary for optimizing CL microscopy, as they influence the resolution, sensitivity, and interpretability of the CL signal.

\begin{figure*}[t]
\includegraphics[width=1\linewidth]{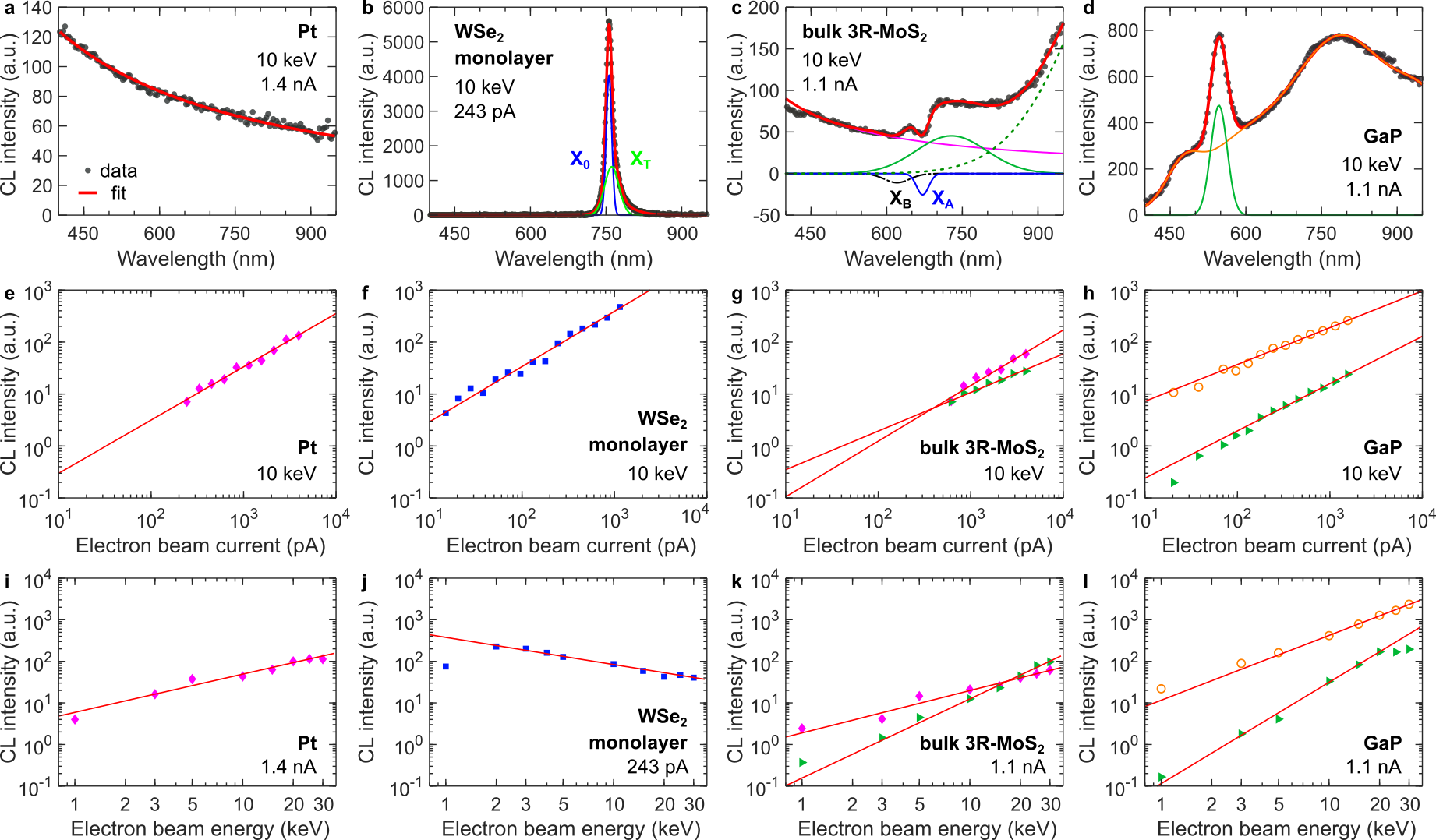}
\caption{\textbf{CL intensity at increasing electron beam current and energy.}	
Spectral fit decomposition of CL from 
\textbf{a}~Pt showing pure transition radiation,
\textbf{b}~hBN-encapsulated WSe$_2$ monolayer where peaks correspond to the neutral exciton $\mathrm{X}_0$ (blue) and trion $\mathrm{X}_\mathrm{T}$ (light green),
\textbf{c}~bulk 3R-MoS$_2$ with mix of transition radiation (magenta), direct (dashed green) and indirect (solid green) bandgap emission, and absorption peaks for $\mathrm{X}_\mathrm{A}$ (solid blue) and $\mathrm{X}_\mathrm{B}$ (dash-dotted black), and
\textbf{d}~GaP with indirect bandgap (green) and defect emission (black). 
The cumulative spectral fits are shown in red.
Evolution of CL intensity with electron beam current (\textbf{e-h}) and voltage (\textbf{i-l}) for \textbf{e,i}~transition radiation in Pt (magenta diamonds), 
\textbf{f,j}~excitonic emission in WSe$_2$ monolayer (blue squares), 
\textbf{g,k}~bulk 3R-MoS$_2$ with mixed transition radiation (magenta diamonds) and bandgap-related emission (green triangles), and \textbf{h,l}~GaP with mixed indirect bandgap (green triangles) and defect emission (orange circles).
The red lines represent power law fits; see text for details.
}
\label{fig-intensity-IV}
\end{figure*}

We investigate the dependence of CL intensity on electron beam current and energy for Pt, an hBN-encapsulated WSe$_2$ monolayer, a bulk 3R-MoS$_2$ crystal, and GaP.
The CL spectra of these materials at 10\,keV are presented in Fig.~\ref{fig-intensity-IV}\pnl{a-d}, using comparable intensity scales.
These materials were selected as they represent the primary sources of CL, including transition radiation, excitonic emission, bandgap-related emission, and defect-related emission.
Fig.~\ref{fig-intensity-IV}\pnl{a} shows a typical broadband CL spectrum of transition radiation measured from bulk Pt, fitted with a function inversely proportional to the emitted wavelength~\cite{Yamamoto1996}, i.e., $I_\mathrm{CL}(\lambda) \sim 1/\lambda$.
This fitting function allows us to extract the CL intensity $I_\mathrm{CL}$ by integrating over the wavelength range from 400\,nm to 950\,nm.
Fig.~\ref{fig-intensity-IV}\pnl{e} shows the evolution of the obtained CL intensity with electron beam current $I$ at a fixed electron beam energy of 10\,keV.
We fit the data using a power law function of the form $I_\mathrm{CL}(I) \sim (I/I_0)^{m_I^\mathrm{TR}}$, where $I_0$ denotes the reference current.
Fig.~\ref{fig-intensity-IV}\pnl{i} presents the results of CL intensity measurements for Pt where the electron beam current is fixed at 1.4\,nA while the electron beam energy $E$ is varied.
Likewise, we fit the obtained CL intensity dynamics using a power law function of the form $I_\mathrm{CL}(E) \sim (E/E_0)^{m_E^\mathrm{TR}}$, where $E_0$ denotes the reference energy.
The CL intensity fits for electron beam current and energy dependencies yield the scaling exponents $m$, summarized for all tested materials in Tab.~\ref{tab:mpower}.
For transition radiation, we observe a linear dependence of CL intensity on both electron beam current and energy, with the scaling exponent $m$ close to 1.

\input{table_m_intensity}

Fig.~\ref{fig-intensity-IV}\pnl{b} presents a narrow CL spectrum for exciton recombination in an hBN-encapsulated WSe$_2$ monolayer on a Si substrate.
THe hBN-WSe$_2$ heterostructure is needed to generate CL because the monolayer is too thin to capture sufficient electron beam energy directly, so the electron beam excites carriers in the wide bandgap hBN, which then recombine through the monolayer to produce CL~\cite{Zheng2017,Fiedler2023tmd}.
To extract the CL intensity, we fit the spectrum with a two-Gaussian function, where the two components represent the neutral exciton and trion contributions to the CL spectrum~\cite{Morozov2024}.
Fig.~\ref{fig-intensity-IV}\pnl{f,j} present the results of CL intensity measurements at fixed electron beam energy and current, respectively.
The power law fit in Fig.~\ref{fig-intensity-IV}\pnl{f} reveals a linear increase in exciton emission intensity with electron beam current.
In contrast, Fig.~\ref{fig-intensity-IV}\pnl{j} shows a decrease in exciton emission intensity with increasing electron beam energy, also following a power law but with a negative exponent.
The excitation of the hBN-WSe$_2$ heterostructure is less efficient at higher electron energies because electrons pass through it without significant interaction and deposit their energy deeper in the Si substrate, where it does not contribute to monolayer excitation.

Fig.~\ref{fig-intensity-IV}\pnl{c} presents the complex CL spectrum of a bulk 3R-MoS$_2$ crystal, similar to that reported in Fig.~\ref{fig-spec-tmds}\pnl{b}.
Angular spectrum measurements, shown in Fig.~\ref{fig-spec-tmds}\pnl{n,p}~{(see SI Fig.~S12)}, confirm that the blue part of the emission spectrum has a coherent origin, while the red part originates from incoherent processes.
Consequently, we attribute the blue part of the CL spectrum to transition radiation, the red part to the direct and indirect bandgap recombination -- both of which can coexist in bulk 3R-MoS$_2$ -- and the spectral dips to absorption by A and B excitons~\cite{Munkhbat2022}. 
Therefore, we fit the complex CL spectrum in Fig.~\ref{fig-intensity-IV}\pnl{c} as a sum of contributions from (i) transition radiation, modeled as $\sim 1/\lambda$ (magenta), (ii) bandgap-related transitions, represented by two Gaussian peaks (solid and dashed green), and (iii) A and B exciton absorption peaks, modeled as negative Gaussians (solid blue and dashed black, respectively).
This spectral decomposition enables us to distinguish between the coherent (transition radiation) and incoherent (bandgap-related emission) contributions to the CL intensity, plotted in Fig.~\ref{fig-intensity-IV}\pnl{g,k} by magenta diamonds and green triangles, respectively.
We apply a power law fit and obtain a linear dependence of transition radiation intensity on the electron beam current and energy in Fig.~\ref{fig-intensity-IV}\pnl{g,k}, which agrees well with the results for bulk Pt. 
The intensity of bandgap-related emission exhibits sub-linear dependence on current and super-linear dependence on energy (Tab.~\ref{tab:mpower}).

Fig.~\ref{fig-intensity-IV}\pnl{d} presents the CL spectrum of GaP, where CL intensity originates from the indirect bandgap emission around 546\,nm and broad defect emission. 
We fit the CL spectrum in Fig.~\ref{fig-intensity-IV}\pnl{d} as the sum of a Gaussian for the indirect bandgap emission (green) and four Gaussians for the broad defect emission (orange). 
Fig.~\ref{fig-intensity-IV}\pnl{h,l} present the results of spectral decomposition highlighting the contributions from indirect bandgap emission (green triangles) and defect emission (orange circles).
The evolution of CL intensity for indirect bandgap and defect emission contributions with increasing electron beam current and energy are fitted with a power law, a sub-linear intensity dependence on current and a super-linear dependence on energy (Tab.~\ref{tab:mpower}).
At higher acceleration voltages, the intensity of indirect bandgap emission saturates, a trend previously observed in doped GaP samples~\cite{Casey1971}.

\begin{figure*}[t]
\includegraphics[width=1\linewidth]{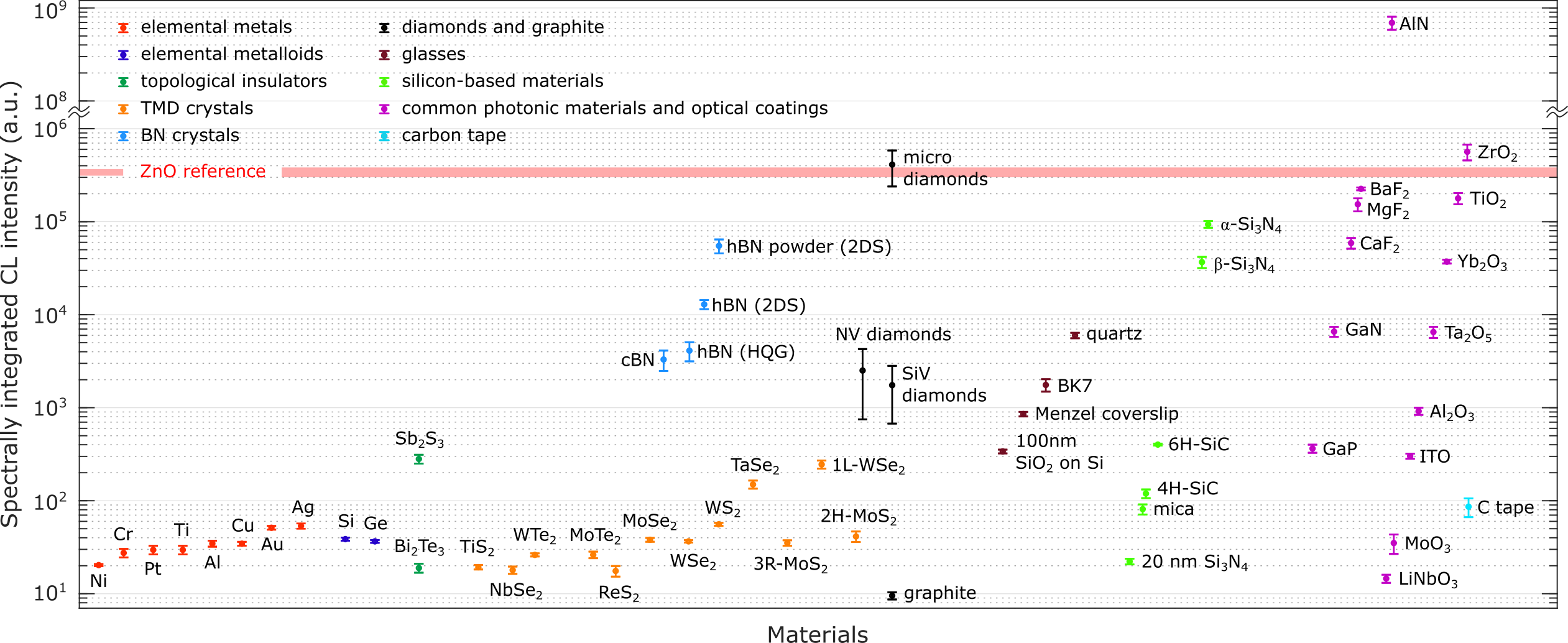}
\caption{\textbf{Comparison of CL intensity for investigated materials.}	
Spectrally integrated CL intensity measured at ca. 1.4\,nA and 10\,keV.
The error bar represents one standard deviation.
The horizontal red-shaded area serves as a reference, representing the response of ZnO powder under identical acquisition and electron beam parameters.
}
\label{fig-intensity-comparison}
\end{figure*}

In summary, the CL intensity of transition radiation shows a clear linear dependence on both electron beam current and energy.
This behavior occurs because transition radiation is a coherent process that scales directly with the number of electrons (beam current) and their energy, as higher-energy electrons generate more photons by more effectively perturbing the electromagnetic field at the material interface.
Similarly, excitonic emission intensity increases linearly with current, as a higher beam current provides more electrons to excite the sample.
However, as electron beam energy increases, excitonic emission intensity in thin heterostructures decreases due to reduced excitation efficiency.
At higher energies, electrons pass through the thin sample with minimal interaction, reducing the excitation efficiency of the hBN-encapsulated monolayer.
In contrast, the CL intensity of bandgap- and defect-related emission exhibits a super-linear increase with electron beam energy, likely due to enhanced excitation efficiency.
Higher-energy electrons penetrate deeper into the material, exciting a larger volume and potentially generating additional photons through secondary processes. 
However, the increased penetration depth could lead to self-absorption effects, resulting in a saturation of the intensity trend.

Finally, we report the spectrally integrated CL intensity for a wide selection of materials under identical electron beam parameters of 1.4\,nA and 10\,kV. 
Each material was measured at five different points on the sample, with the setup realigned from scratch each time, allowing us to determine the mean CL intensity and its standard deviation, shown as error bars in Fig.~\ref{fig-intensity-comparison}. 
As a reference, we used a ZnO powder, which commonly employed for alignment and calibration of CL systems, with its intensity indicated as a red-shaded area in Fig.~\ref{fig-intensity-comparison}.
Additionally, we report CL intensity for typical carbon tape used in CL microscopy. 
Reliable CL intensity values could not be obtained for polymeric and resist materials due to the specific CL measurement procedure and their rapid degradation under electron beam exposure.

\section{7 Conclusions}
This work presents a comprehensive atlas of CL response for a broad range of materials relevant to photonic and plasmonic applications. 
Through Monte Carlo simulations, we demonstrate how variations in electron beam energy and material density affect electron penetration and energy deposition profiles.
By systematically characterizing materials under electron beam excitation across a range of energies, we provide detailed insights into the origins of coherent and incoherent CL. 
Our results highlight the distinct CL properties of metals, metalloids, topological insulators, TMDs, boron nitride, carbon-based materials, commonly used photonic materials and optical coatings, and polymers.
Our findings aid in the selection of materials and substrates for advanced photonic and plasmonic experiments in an electron microscope.
Additionally, this atlas serves as a foundational resource for future CL studies, offering guidance on experimental parameters for optimal material characterization.
This resource ultimately paves the way for novel applications in photonics and plasmonics, fostering innovative research in these fields.\\

\noindent
\textbf{Acknowledgments:} The Center for Polariton-driven Light--Matter Interactions (POLIMA) is sponsored by the Danish National Research Foundation (Project No.~DNRF165).
We thank Dr.~Catarina Ferreira for her comments and suggestions during the revision of the manuscript.

\noindent
\textbf{Author contributions:} The idea of an atlas was conceived by S.~E. and S.~M., who, along with Y.~L. and T.~Y., contributed to the sample preparations. 
Cathodoluminescence spectroscopy was carried out by S.~E. and S.~M., while Y.~L. and T.~Y. performed Raman spectroscopy. 
The project was supervised by S.~M. and N.~A.~M.
All authors contributed to analyzing the data and writing the manuscript. 
All authors have accepted responsibility for the entire content of this manuscript and approved its submission.

\noindent
\textbf{Data availability:} The datasets generated and/or analyzed during the current study are available from the corresponding author upon reasonable request.

\bibliography{bibliography}
\bibliographystyle{unsrt}

\end{document}

%% file: table_m_intensity.tex
\setlength{\extrarowheight}{4pt} 

\begin{table}[]
\begin{tabular}{|l|c|c|c|c|c|c|c|c|}
\hline
Material &
  $m^\mathrm{TR }_\mathrm{I}$ &
  $m^\mathrm{TR}_\mathrm{E}$ &
  $m^\mathrm{X}_\mathrm{I}$ &
  $m^\mathrm{X}_\mathrm{E}$ &
  $m^\mathrm{BG}_\mathrm{I}$ &
  $m^\mathrm{BG}_\mathrm{E}$ &
  $m^\mathrm{D}_\mathrm{I}$ &
  $m^\mathrm{D}_\mathrm{E}$ \\ 
  \toprule 
  \hline
Pt                                                         &1.02&0.92&--  &--   &--  &--  &--  &--\\ \hline
\begin{tabular}[c]{@{}l@{}}WSe$_2$\\ monolayer\end{tabular} &--  &--  &1.06&-0.65&--  &--  &--  &--\\ \hline
\begin{tabular}[c]{@{}l@{}}bulk\\ 3R-MoS$_2$\end{tabular}  &0.96&1.06&--  &--   &0.70&1.90&--  &--\\ \hline
GaP                                                        &--  &--  &--  &--   &0.91&2.44&0.71&1.56\\ \hline
\end{tabular}
\caption{Power law fitting of CL intensity evolution at increasing electron beam current ($m_\mathrm{I}$) and energy ($m_\mathrm{E}$). The superscript in scaling exponent $m$ stands for emission of transition radiation (TR), excitons~(X), bandgap-related emission~(BG), and defects~(D). }
\label{tab:mpower}
\end{table}